\documentclass[11pt,preprint]{aastex}
\usepackage{amsmath}
\usepackage{graphicx}
\graphicspath{{/Users/casan/files/publi/papier_local_emissivity/arXiv}}
\usepackage{ulem}
\usepackage{natbib}
\usepackage{lineno}


\begin{document}
\modulolinenumbers[1]

\title{Local H~{\sc i} emissivity measured with the {\it Fermi}-LAT and implications for cosmic-ray spectra}
\author{Jean-Marc Casandjian}
\affil{Laboratoire AIM, CEA-IRFU/CNRS/Universit\'e Paris Diderot, Service d'Astrophysique, CEA Saclay, 91191 Gif sur Yvette, France}
\email{casandjian@cea.fr}

\begin{abstract}

Cosmic-ray (CR) electrons and nuclei interact with the Galactic interstellar gas and produce high-energy $\gamma$ rays. The $\gamma$-ray emission rate per hydrogen atom, called emissivity, provides a unique indirect probe of the CR flux. We present the measurement and the interpretation of the emissivity in the solar neighborhood for $\gamma$-ray energy from 50~MeV to 50~GeV. We analyzed a subset of 4 years of observations from the Large Area Telescope (LAT) aboard the {\it Fermi Gamma-ray Space Telescope} ({\it Fermi}) restricted to absolute latitudes $10\degr<\left| b \right| <70\degr$. From a fit to the LAT data including atomic, molecular and ionized hydrogen column density templates as well as a dust optical depth map we derived the emissivities, the molecular hydrogen to CO conversion factor $X_{CO}=(0.902\pm0.007) \times 10^{20}$ cm$^{-2}$ (K km s$^{-1}$)$^{-1}$ and the dust-to-gas ratio $X_{DUST}=(41.4\pm0.3) \times 10^{20}$ cm$^{-2}$ mag$^{-1}$. Moreover we detected for the first time $\gamma$-ray emission from ionized hydrogen. We compared the extracted emissivities to those calculated from $\gamma$-ray production cross-sections and to CR spectra measured in the heliosphere. We observed that the experimental emissivities are reproduced only if the solar modulation is accounted for. This provides a direct detection of solar modulation observed previously through the anticorrelation between CR fluxes and solar activity. Finally we fitted a parametrized spectral form to the heliospheric CR observations as well as to the {\it Fermi}-LAT emissivity and obtained compatible local interstellar spectra for proton and Helium kinetic energy per nucleon between between 1 and 100~GeV and for electron-positrons between 0.1 and 100~GeV. 
\end{abstract}

\maketitle

\section{Introduction}

The theory behind cosmic rays and interstellar medium (ISM) interactions developed in the 1950s \citep{Hayakawa:1952p3259, Hutchinson:1952p3262, Ginzburg:1965p3189, Stecker:1970p3081, Stecker:1971p3052} was first experimentally addressed with balloons and rockets experiments in the sixties. However the first evidence of correlation between high-energy $\gamma$ rays and Galactic structures was observed with the launch of the Third Orbiting Solar Observatory OSO-3 in March 1967 \citep{Kraushaar:1972p3349}. A $\gamma$-ray H~{\sc i} emissivity of 1.6$\times$10$^{-25}$ ph atom$^{-1}$ s$^{-1}$ above 100~MeV was derived from comparison with an early H~{\sc i} 21-cm line radiation radio survey. The Small Astronomy Satellite 2 (SAS-2) \citep{Fichtel:1975p3076} launched in November 1972 collected 20 times more photons than OSO-3 and provided clearer evidence of correlation between high-energy $\gamma$ rays and hydrogen \citep{Kniffen:1977p3084}. \cite{Lebrun:1979p3085} compared the diffuse gamma-ray intensities measured by SAS-2 with atomic hydrogen (H~{\sc i}) column density and derived an integrated H~{\sc i} $\gamma$-ray emissivity  above 100~MeV of  (2.9$\pm$0.6)$\times$10 $^{-25}$ ph atom$^{-1}$ s$^{-1}$. Later \cite{Lebrun:1982p3092} extracted  the $\gamma$-ray emissivity of H~{\sc i} in three energy bands from the Cosmic Ray Satellite (COS-B) observations \citep{Bennett:1976p3256}. They assumed a power-law spectrum for the $\gamma$ rays and obtained a differential emissivity of 2.1 $\times$ 10$^{-23}$ $\times$ E$^{-2}$ ph atom$^{-1}$ s$^{-1}$ MeV$^{-1}$ and an integrated flux above 100~MeV of 2.1$\pm$ 0.3$\times$ 10$^{-25}$ ph atom$^{-1}$ s$^{-1}$. Using the same data and a model for the interstellar emission that included $\gamma$ rays also originating from the molecular hydrogen and inverse-Compton scattering (IC), \cite{Strong:1988p3093} investigated the radial variation of the emissivity and the systematic uncertainties associated to the model. The Energetic Gamma Ray Experiment Telescope (EGRET) \citep{Thompson:1993p247} on the Compton Gamma-Ray Observatory, launched in 1991, provided a higher angular and energy resolution and collected about 4 times more  $\gamma$ rays than COS-B above 100~MeV. In \cite{Strong:1996p1032} an all-sky accurate interstellar model and a spectrum of the H~{\sc i} $\gamma$-ray emissivity were presented, followed by various publications related to the interpretation of an apparent excess observed around 1~GeV in the EGRET data.

The Large Area Telescope (LAT) is the main instrument onboard the {\it Fermi} satellite. It collects $\gamma$ rays in the energy range 20~MeV to greater than 300~GeV. The LAT has a wide field of view of about 2.4~sr, an on-axis effective area of ∼8000~cm$^2$ and a 68\% containment of the point-spread function (PSF) of 0.8$\degr$ at 1~GeV. For most of the first four years of the mission it was operated in all-sky survey mode and imaged the sky every two orbits in 3~hr. \cite{Abdo:2009p3049} used LAT photons in a mid-latitude region in the third quadrant lacking known molecular clouds to extract the $\gamma$-ray emissivity per H~{\sc i} in the neighborhood of the solar system for energies from 100~MeV to 9~GeV. They masked the bright $\gamma$-ray point sources, and correlated for each energy bin the H~{\sc i} column density ($N$(H~{\sc i}) ) with the LAT counts after subtracting the contribution from IC. With the post-launch response function Pass 6$\_$V3 DIFFUSE, they measured an integrated $\gamma$-ray emissivity of (1.63$\pm$0.05)$\times$10$^{-26}$ ph s$^{-1}$sr$^{-1}$atom$^{-1}$ above 100~MeV and found that the differential emissivity agrees with calculations based on CR spectra consistent with those directly measured. The EGRET apparent excess not confirmed by {\it Fermi} \citep{Abdo:2009p4020} was of likely instrumental origin \citep{Stecker:2008p4103}. \cite{Delahaye:2011p2770} compared those emissivities with predictions from various proton and Helium local interstellar spectra (LIS) and a $\gamma$-ray production cross-section of \cite{Huang:2007p2906}. They found good agreement with the LAT measurements but neglected the contribution from electrons. Emissivity was also derived from {\it Fermi}-LAT observations in specific regions like Cassiopeia and Cepheus \citep{Abdo:2010p3307}, the third Galactic quadrant \citep{Ackermann:2011p4125}, Orion \citep{Ackermann:2012p4114}, the Cygnus region \citep{Ackermann:2012p4119} and Chamaeleon, R Coronae Australis and Cepheus and Polaris flare regions \citep{Ackermann:2012p4116,Ackermann:2013p4117}.

The objective of the present work is to derive the spectra of local interstellar CRs from {\it Fermi}-LAT observations independent of solar modulation which affects direct measurements in the heliosphere. We have extended the work of \cite{Abdo:2009p3049} to the whole longitude range and to latitude $10\degr<\left| b \right| <70\degr$ to derive from 4 years of LAT data the local H~{\sc i} emissivity using a template fitting method. This method is based on the fitting of the {\it Fermi}-LAT counts map by a linear combination of templates spatially correlated with predicted production sites of $\gamma$ rays. We compare those emissivities with the ones computed from heliospheric CR observations and $\gamma$-ray production cross-sections. We derive an electron-positron, proton and Helium LIS compatible with the H~{\sc i} emissivity using the proton and Helium fluxes detected with PAMELA \citep{Adriani:2011p4018} as well as the electron plus positron fluxes measured by {\it Fermi}-LAT \citep{Ackermann:2010p4023}.

\section{ISM Neutral Gas Census}
Most $\gamma$ rays in the LAT energy range produced in the ISM come from CR protons impinging on atomic and molecular hydrogen and, to a lesser extent for the energies considered here, bremsstrahlung radiation when electrons and positrons are deflected by the gas nuclei. The number of $\gamma$ rays produced is then expected to be proportional to the neutral hydrogen column density. CRs also interact with the Helium and heavier-element gases of the ISM, here we assume that they are uniformly mixed with the hydrogen. We used the all-sky Leiden-Argentine-Bonn (LAB) \citep{Kalberla:2005p3048} 21-cm line radiation composite survey to obtain the atomic hydrogen $N$(H~{\sc i}) column densities with the assumption of a uniform spin temperature ($T_{S}$). We used the J=1$\rightarrow$0 line of carbon 12 monoxide (CO) at 2.6-mm as a tracer for molecular hydrogen H$_{2}$. The velocity integrated CO intensity $W$(CO) was obtained from the Center for Astrophysics compilation \citep{Dame:2001p1849} as well as new observations extending the coverage in the northern hemisphere with the same telescope (T. Dame 2011, private communication). In our study, we did not assume a value for the molecular hydrogen to CO conversion factor $X_{CO}= N({\rm H_2})/W({\rm CO})$; instead, we obtained it from the fitting procedure described in Section ~\ref{sec:Fitting_procedure_section}. The "Dark Neutral Medium" (DNM) gas \citep{Grenier:2005p836,Ade:2011p3178,Paradis:2012p4067}, found in excess over estimations based on $N$(H~{\sc i}) and $W$(CO), was accounted for by including in our model a template for total dust column-density since neutral gas and dust column-density are to the first order linearly related \citep{Boulanger:1996p2089}. We used the dust optical depth map of \cite{Schlegel:1998p290} to trace the total dust column density. This map is based on the IRAS/ISSA map at 100~$\micron$ with the infrared intensity corrected for dust temperature based on the ratio of COBE/DIRBE at 100 and 240 $\micron$. It was normalized as E(B--V) reddening expressed in mag. We modeled the dust optical depth map with a linear combination of $N$(H~{\sc i}), $W$(CO) and isotropic emission maps with coefficients free to vary. We fitted this simple model to the dust column-density map and obtained a residual dust map corresponding to dust not linearly correlated with the $N$(H~{\sc i}) and $W$(CO) column-densities. The final DNM map was obtained after applying to this residual map an absolute magnitude cut at 0.04~mag to remove diffuse high latitude dust residual clouds in low density regions where dust optical depth and hydrogen column density depart from portionality (see Figure 6 of \cite{Ade:2011p3178}). 

To allow for a Galactocentric gradient of CR flux in the Galaxy, we partitioned the $N$(H~{\sc i}) column-densities and W(CO) intensities into several Galactocentric annuli. The present work interprets the emissivities of the local H~{\sc i} annuli corresponding to Galactocentric distances from about 8 to 9.5~kpc (the IAU recommended value for the Sun Galactocentric distance is 8.5~kpc). Restricting the analysis to absolute latitude above 10$\degr$ insured that most $\gamma$ rays originated from collisions between CR and hydrogen located at distances within few hundreds of parsecs from the sun. A more precise description of the gas annuli used in this work is given in \cite{Ackermann:2012p2978}.

\section{Ionized Hydrogen}
\label{sec:Ionized_hydrogen}
Hydrogen also exists in form of H~{\sc ii} regions embedded in a pervasive diffuse plasma layer \citep{Reynolds:1991p4085,Ferriere:2001p1346,Gaensler:2008p3776,Haffner:2009p4086}. This layer called the warm ionized medium (WIM), or sometimes ``Reynolds Layer'', represents nearly 90\% of the H$^{+}$ with a scale height of 8 times the scale height of the neutral hydrogen \citep{Gaensler:2008p3776}. First introduced to interpret the low-frequency absorption of Galactic synchrotron, it was later confirmed based on the dispersion measures of pulsars and detected through optical recombination line emission (mainly H$\alpha$) and Coulomb interactions of free electrons with ions (thermal bremsstrahlung or free-free radiation) at microwave frequency. 

The total mass of H$^{+}$ in the Galaxy is estimated from the combination of dispersion measures of pulsars and models of the density distribution of Galactic free electrons. The model NE2001 developed by \cite{Cordes:2002p3302} is commonly used in pulsar studies, \cite{Schnitzeler:2012p4008} compared dispersion measures to predictions from eight electron density models. The conclusion was that a modified version of NE2001 best reproduces the observations. In NE2001 the WIM is modeled by a thick disk with a squared hyperbolic secant dependence on the height above the Galactic disk. Additionally the model includes a thin annular disk, spiral arms, local bubbles and individual regions of high and low electron density (clumps and voids). The model parameters are tuned to best reproduce 1143 dispersion measures of pulsars with known distances. Since the Helium is neutral in the WIM and since its contribution to the hot ionized medium electron density is limited, we calculated the ionized hydrogen density $N$(H$^{+}$) directly as the free electron density given by NE2001. We obtained from NE2001 that $N$(H$^{+}$)/$N$(H~{\sc i}) equals 0.17 in the Galactic plane $|b|<10\degr$ and 0.3 outside this region. Alternatively, \cite{He:2013p4087} studied 68 radio pulsars detected at X-ray energies and compared the free electron column density given by dispersion measures to $N$(H~{\sc i}) along the line of sight traced by X-ray extinction, they obtained a ratio in the range 0.07 to 0.14.

As we describe below, we fitted various $\gamma$-ray production site templates and point sources to the LAT data (see Section \ref{sec:Fitting_procedure_section}). We added to the fit a template of the WIM based on NE2001 predictions with a normalization factor free to vary, this component did not improve the model likelihood. We did not observe any improvement when applying latitude cuts or by removing individual H$^{+}$ clumps that seemed over predicted. \cite{Paladini:2007p813} also observed a dust emission fit worsen with an H$^{+}$ template extracted from NE2001. This probably comes from the simplified model used in NE2001, Figure 7 of \cite{Sun:2008p3774} shows that NE2001 does not reproduce the structures of the free-free emission template from Wilkinson Microwave Anisotropy Probe (WMAP) observations. Based on \cite{Gaensler:2008p3776} we also created an H$^{+}$ template based on an exponential scale-height for H$^{+}$ leading to a cosecant $N$(H$^{+}$) template map which also did not improve the model likelihood in the fit to the LAT observations. 

While the dispersion measure provides the column density of free electrons, the free-free and H$\alpha$ emissions, which involve two-particle processes, are proportional to the integral of the square of the electron density along the line of sight (emission measure). H$\alpha$ emission suffers from dust absorption particularly for absolute latitude below 15$\degr$ \citep{Dickinson:2003p4084} and a substantial fraction of the diffuse high-latitude H$\alpha$ intensity is the result of scattering by interstellar dust of H$\alpha$ originating elsewhere in the Galaxy \citep{Witt:2010p4082,Seon:2012p4083}. The free-free emission does not suffer from those flaws but microwave surveys also include synchrotron radiations. We used the free-free intensity map at frequency $\nu$=22.7~GHz ($I_{ff}$) extracted from 9-years of WMAP observations with the maximum entropy method \citep{Bennett:2013p4019} as a proxy for the $\gamma$-ray emission correlated with H$^{+}$. 

The conversion from emission measure to column density is done by assuming an effective electron density $n_{eff}$ such that the integral of the square of the electron density $n_{e}$ along the line of sight is $\int n_{e}^{2}ds = n_{eff} \int n_{e}ds$. Under this approximation, the free-free emission intensity in Jy sr$^{-1}$ is $I_{ff}=2.5\times10^{6}{T_{e}}^{-0.35}\nu^{-0.1}n_{eff}\int{n_{e}ds}$ where the electron temperature $T_{e}$ is in K, the frequency $\nu$ in GHz and the line of sight distance in kpc \citep{Sodroski:1997p4010}. We can then express the H$^{+}$ column density as: $N$(H$^{+}$)=$\int{n_{e}ds}=1.2\times10^{15}{T_{e}}^{0.35}\nu^{0.1} {n_{eff}}^{-1} I_{ff}$. We used $N$(H$^{+}$) in Equation \ref{eqRing} adopting an electron temperature $T_{e}$=8000~K \citep{Sodroski:1989p4012}.

\section{Inverse Compton and Large Scale Structures}

The inverse Compton scattering of the interstellar radiation field by electrons and positrons represents a non-negligible fraction of the $\gamma$-ray interstellar emission \citep{Ackermann:2012p2978}. There is no simple template for the IC emission so we relied for its morphology on the program GALPROP\footnote{\url{http://galprop.stanford.edu}} \citep{Strong:2007p1625,Vladimirov:2011p4099} which predicts the CR density by solving numerically the transport equation of charged particles, starting from a given source distribution and propagation parameters. This density is combined by GALPROP with an interstellar radiation field model \citep{Porter:2008p3784} into a predicted IC intensity map $I_{IC_{p}}$. We used an IC morphology predicted by one of the diffusive reacceleration models described in \cite{Ackermann:2012p2978} with the identification $^SY^Z6^R30^T150^C2$ (CR source distribution proportional to the radial distribution of pulsars in the Galaxy given by \cite{Yusifov:2004p3648}, Galactic halo size and radial boundary equal to 6~kpc and 30~kpc). In Section \ref{sec:systematic_errors} we also used two different IC templates also derived from GALPROP diffusive reacceleration models but with different input parameters designated 54\_z10G4c5rS (Gaussian source distribution with a halo height of 10~kpc) and 54\_77Xvarh7S (source distribution intermediate between the radial distribution of SNRs and that of pulsars, halo height of 4~kpc). During the fitting procedure, we left the $I_{IC_{p}}$ normalization free in each energy bin to allow some freedom for possible variations of the CR distribution and Galactic interstellar radiation field from the GALPROP model.

GALPROP does not include local structures like the Loop~{\sc i} giant radio continuum loop. Gamma rays likely coming from IC emission of CR electrons trapped in Loop~{\sc i} have been reported \citep{Grenier:2005p4078,Casandjian:2009p3311,Ackermann:2012p2978}. We modeled this emission using the 408~MHz radio-continuum emission survey ($I_{LoopI}$) of \cite{Haslam:1982p4092}. We subtracted the radio point sources from this map and created a dedicated template from a selected region around Loop~{\sc i}. Since the 408~MHz radio-continuum emission depends on the magnetic field while the $\gamma$ emission does not, we cannot rely only on the radio template for our analysis. That is, we cannot assume that the $\gamma$-ray intensity is proportional to the radio continuum intensity. We extended the radio template of Loop~{\sc i} with a patch of uniform intensity ($I_{patche}$) with corners located at $(l,b)= (350\degr,0\degr), (310\degr,25\degr), (300\degr,17\degr), (332\degr,-12\degr)$. This shape approximately reproduces the extension observed map of residuals between the LAT observations and the model when only the 408~MHz radio-continuum template was used, this patch possibly compensates for the difference between the radio and $\gamma$-ray emission of Loop~{\sc i} in a region where the magnetic field might be weaker.

Additionally we have incorporated in our model a template for the {\it Fermi} Bubbles \citep{Su:2010p2675,Ackermann:2014p4189}. There is no accurate a priori template for the $\gamma$-ray emission of those large structures, but their morphologies are distinct from those of the gas or of the GALPROP IC templates and their spectra are harder. We derived their shapes from an iterative procedure for energies integrated above 4~GeV, starting from a fit using a uniform intensity template with the same shape as that of the Bubbles. We added to this template the fit residuals and verified that the template derived above 4~GeV ($I_{Bubbles}$) gives a flat residual map at lower energies.

\section{Data Selection}
\begin{figure}
\begin{center}
\includegraphics[width=14cm]{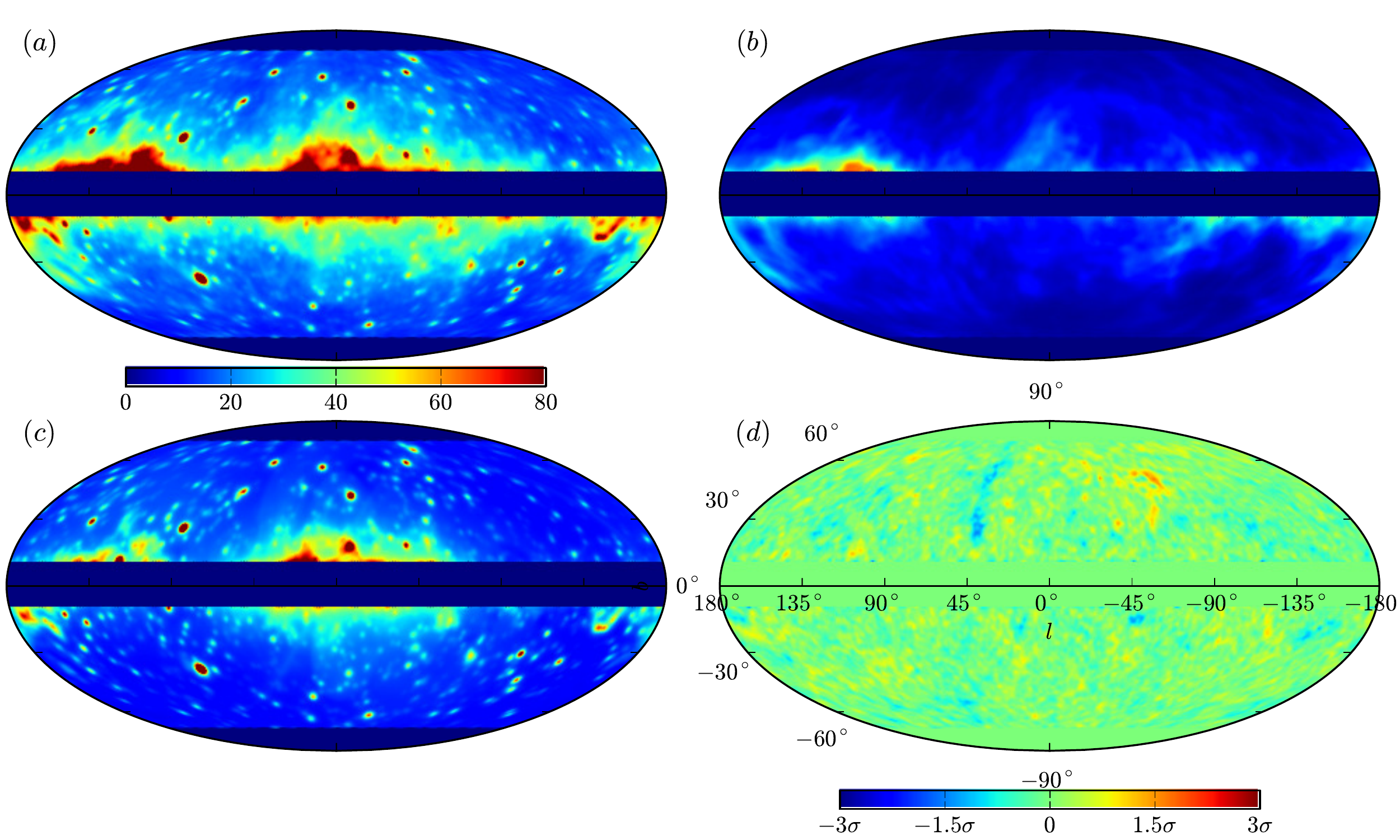}
\caption{a: {\it Fermi}-LAT counts per pixel integrated above 170~MeV together with predictions calculated from Equation \ref{eqRing} for $\gamma$ rays originating from local H~{\sc i} (b) and from remaining contributions (c). When the contributions from (b) and (c) are subtracted from the total counts map (a) we obtain the residual plot (d) displayed in units of standard deviations. We used the same scale for maps (a--c).  We masked the absolute latitudes below 10$\degr$ and above 70$\degr$ corresponding to region outside our region of interest. In this figure the maps are displayed in Mollweide projection in Galactic coordinates, we applied a smoothing with a gaussian symmetric beam of 2$\degr$ (FWHM). The pixel size for all the maps is 0.25$\degr$.}
\label{LAT_counts}
\end{center}
\end{figure}

Our analysis is based on Pass 7 reprocessed data (P7REP) Clean class events for the first 4 years of the mission. We excluded from this time interval periods during intense transients (fluent GRBs, bright AGN or solar flares) occurred \citep{Ballet:2013p4107}. We used the Clean selection that has a reduced residual charged-particle background compared to the Source selection \citep{Ackermann:2012p4094}. The event selection included 55 million LAT counts which were binned into 26 energy intervals logarithmically spaced from 50~MeV to 50~GeV. For energies below 50~MeV, at which the Pass 7 effective area is approximately 20 times lower than at 1~GeV, we did not obtain a good fit stability. The lower statistics, a higher Earth limb contamination and a broadening of the LAT point-spread function (PSF) led to confusion between point sources and Galactic interstellar emission. The highest four energy bins were combined into two bins. In each energy interval the high statistics allowed a good separation of photons correlated with H~{\sc i} from those with other origins for absolute latitude above 10$\degr$. Figure \ref{LAT_counts} (a) shows the total {\it Fermi}-LAT 4-year accumulated number of counts above 170~MeV ($N_{obs}$) for the latitude range considered here.

The residual intensity from Earth limb photons entering the LAT at large zenith angles ($I_{limb}$) cannot be neglected in the lowest energy bins. It is structured around the celestial poles and can bias the fit. We derived a template for this residual contamination of limb photons by subtracting a counts map derived with a zenith angle cut at 100$\degr$ to one restricted at angles above 80$\degr$. We added this contribution to the model as a template free to vary.

\section{Fitting Procedure}
\label{sec:Fitting_procedure_section}
The template method, based on fitting the $\gamma$-ray observations with a linear combination of maps spatially correlated with production sites of $\gamma$ rays (templates), has been widely used in diffuse $\gamma$-ray astronomy \citep[e.g.,][]{Lebrun:1979p3085,Strong:1988p3093,Digel:1999p335,Grenier:2005p836,Abdo:2010p3307}. The fitting procedure is similar to the one used to build the {\it Fermi}-LAT interstellar emission model \footnote{\url{http://fermi.gsfc.nasa.gov/ssc/data/access/lat/Model\_details/FSSC\_model\_diffus\_reprocessed\_v12.pdf}}. We modeled the predicted spatial distribution of counts ($N_{pred}$) detected by {\it Fermi}-LAT as:

{\footnotesize
\begin{equation}
\begin{split}
N_{pred}(E,l,b) &= \sum_{i=gas} q_{i}(E)\widetilde{I}_{i}(l,b)  +  N_{IC}(E)\widetilde{I}_{IC_{p}}(E,l,b) +  N_{iso}(E)\widetilde{I}_{iso}   + N_{LoopI}(E)\widetilde{I}_{LoopI}(l,b)    \\ & + N_{patch}(E)\widetilde{I}_{patch}(l,b)   + N_{Bubbles}(E)\widetilde{I}_{Bubbles}(l,b) + \sum_{i=point~src} N_{pt_{i}}(E)\widetilde{\delta}(l,b,i)  \\ & + 
\sum_{i=extend~src} N_{ext_{i}}(E)\widetilde{I}_{i}(l,b)  +  \widetilde{I}_{sun\_moon}(E,l,b) + N_{limb}(E)\widetilde{I}_{limb}(l,b) 
\end{split}
\label{eqRing}
\end{equation}
}

where $E$ is the energy and $l$,$b$ are Galactic sky coordinates. 
In Equation \ref{eqRing} the term ``gas'' refers to $N$(H~{\sc i}), $W$(CO), $N$(H$^{+}$) and the DNM column densities.  They are multiplied by the respective hydrogen $\gamma$-ray emissivities $q_{i}$. 
The factors $N_{IC}$, $N_{iso}$, $N_{LoopI}$, $N_{patch}$, $N_{Bubbles}$ and $N_{limb}$ represent the renormalization factors associated with the predicted IC, the isotropic emission represented by a uniform intensity template, the Loop~{\sc i} radio emission and patch, the bubbles and the limb template maps.
The normalization of the templates\footnote{http://fermi.gsfc.nasa.gov/ssc/data/access/lat/2yr\_catalog/} corresponding to the extended emission of the two Magellanic Clouds and of Centaurus A are written $N_{ext_{i}}$. The 2179 point sources listed in a preliminary version of the 3FGL catalog \citep{Ballet:2013p4107} are also fitted individually in Equation \ref{eqRing}, it is represented by a number of counts $N_{pt_{i}}$ and a Dirac $\delta$ function at the source position.
In Equation \ref{eqRing} we denote by $\widetilde{I}$ the counts map resulting from the convolution with the LAT PSF of the product of an intensity map $I$ by the instrument exposure and pixel solid angle. We obtained the gas emissivities and normalization factors from a binned maximum-likelihood fit with Poisson statistics to the {\it Fermi}-LAT counts map in the 26 energy intervals \citep{Mattox:1996p495}. We used the code MINUIT\footnote{\url https://wwwasdoc.web.cern.ch/wwwasdoc/minuit/minmain.html} to calculate the likelihood maxima as well as the parameters error. The intensity from the Sun and the Moon \citep{Abdo:2011p4076,Abdo:2012p4077}, ${I}_{sun\_moon}$, is kept fixed in the analysis.
We restricted the fit to latitudes $10\degr<\left| b \right| <70\degr$ but included in Equation \ref{eqRing} lower-latitude sources and gas templates to account for the spread of the bright Galactic ridge above $10\degr$ at low energy. We excluded regions with absolute latitudes above 70$\degr$ where the isotropic emission dominates. We binned all the maps in HEALPix\footnote{\url{http://healpix.sourceforge.net}} with a Nside of 256 which corresponds to a bin size of about $0\fdg25 \times 0\fdg25$. In each energy interval $q_{i}$, $N_{IC}$, $N_{iso}$, $N_{LoopI}$, $N_{patch}$, $N_{Bubbles}$, $N_{pt_{i}}$, $N_{ext_{i}}$, and $N_{limb}$ were left free to vary.

\section{Results}
\begin{figure}
\begin{center}
\includegraphics[width=14cm]{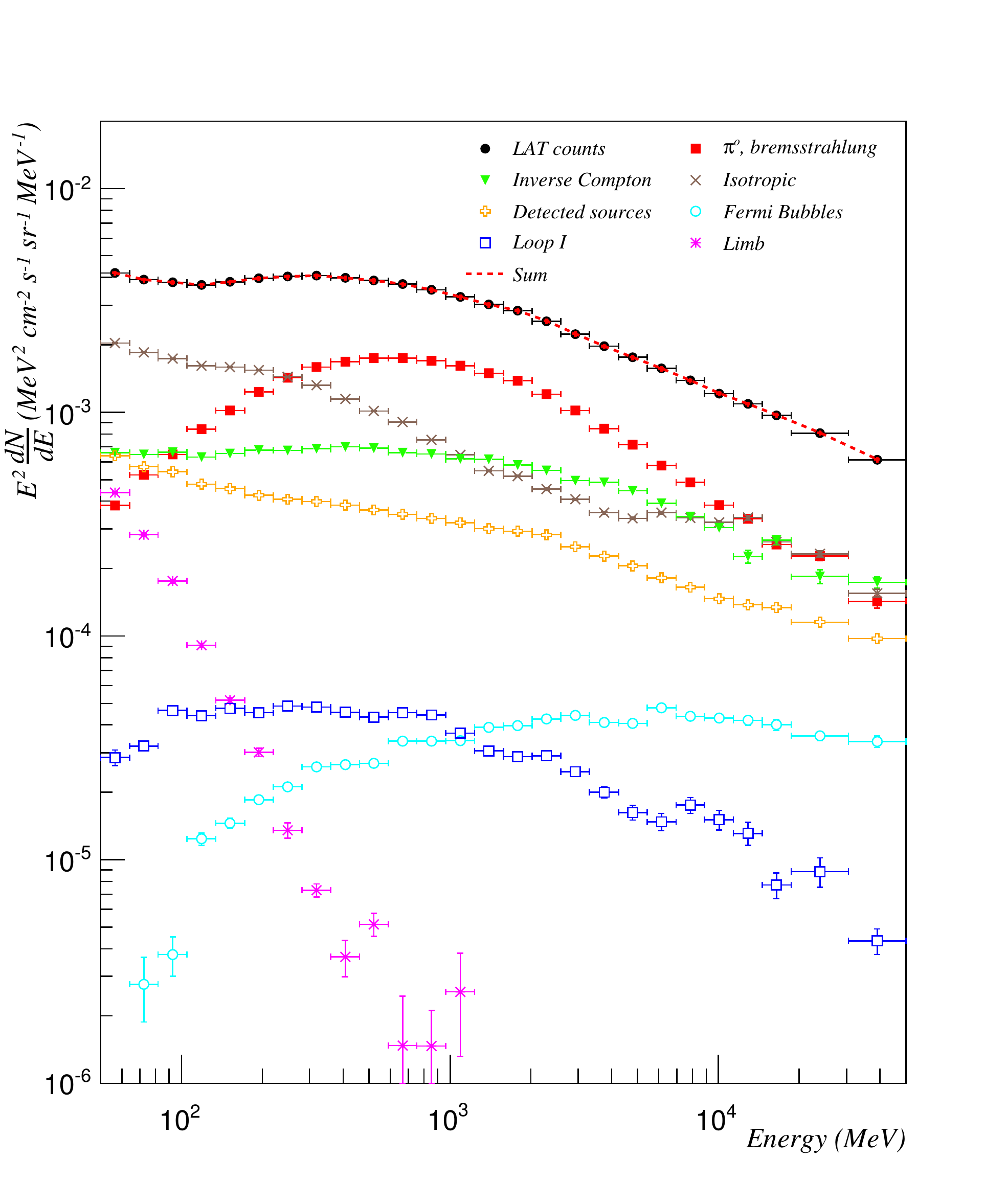}
\caption{Intensity spectrum for the LAT counts and individual contributions of Equation \ref{eqRing} for latitudes $10\degr<\left| b \right| <70\degr$. The uncertainties associated with the LAT intensity are only statistical. The vertical error bars associated with each contribution represent the uncertainty obtained from the fit. For the most part they are smaller than the symbol size. The contributions associated with hydrogen in atomic, molecular, ionized, or DNM phase are merged into a single ``$\pi^{o}$, bremsstrahlung'' contribution.}
\label{spectrum}
\end{center}
\end{figure}

We fitted the model given by Equation \ref{eqRing} to the {\it Fermi}-LAT counts in each of the 26 energy bins. We derived for each template and source an intensity that we averaged over the region studied here and plotted in Figure \ref{spectrum} with vertical error bars obtained from the likelihood maximization. In this graph we also show the intensity calculated from the LAT counts for the same region with the corresponding statistical uncertainties. Horizontal error bars represent the bin energy widths. To simplify the graph we combined the various gas contributions into a single spectrum called ``$\pi^{o}$, bremsstrahlung''; for this spectrum we combined quadratically the uncertainties of individual contributions. We observe a very good agreement between the LAT observations and the sum of the template model intensities. We note that $\gamma$ rays originating from the interaction between the CRs and the gas dominate for energies between 200~MeV and 10~GeV.

We study and interpret in this paper the contribution in Equation \ref{eqRing} from the atomic hydrogen template: $q_{HI} \widetilde{N}$(H~{\sc i}). In Figure \ref{LAT_counts}b we show this contribution, in counts per pixel integrated above 170~MeV. In Figure \ref{LAT_counts}c we show the complementary contribution from all the other terms of Equation \ref{eqRing}: $N_{pred} - q_{HI} \widetilde{N}$(H~{\sc i}). When subtracting all the contributions to the LAT counts map we obtained a flat residual map  $(N_{obs}-N_{pred})/\sqrt{N_{pred}}$ (Figure \ref{LAT_counts}d) which indicates that the main LAT $\gamma$-ray sources are included in Equation \ref{eqRing} and then that the local H~{\sc i} template does not compensate for any un-modelled emission. 

\begin{figure}
\begin{center}
\includegraphics[width=14cm]{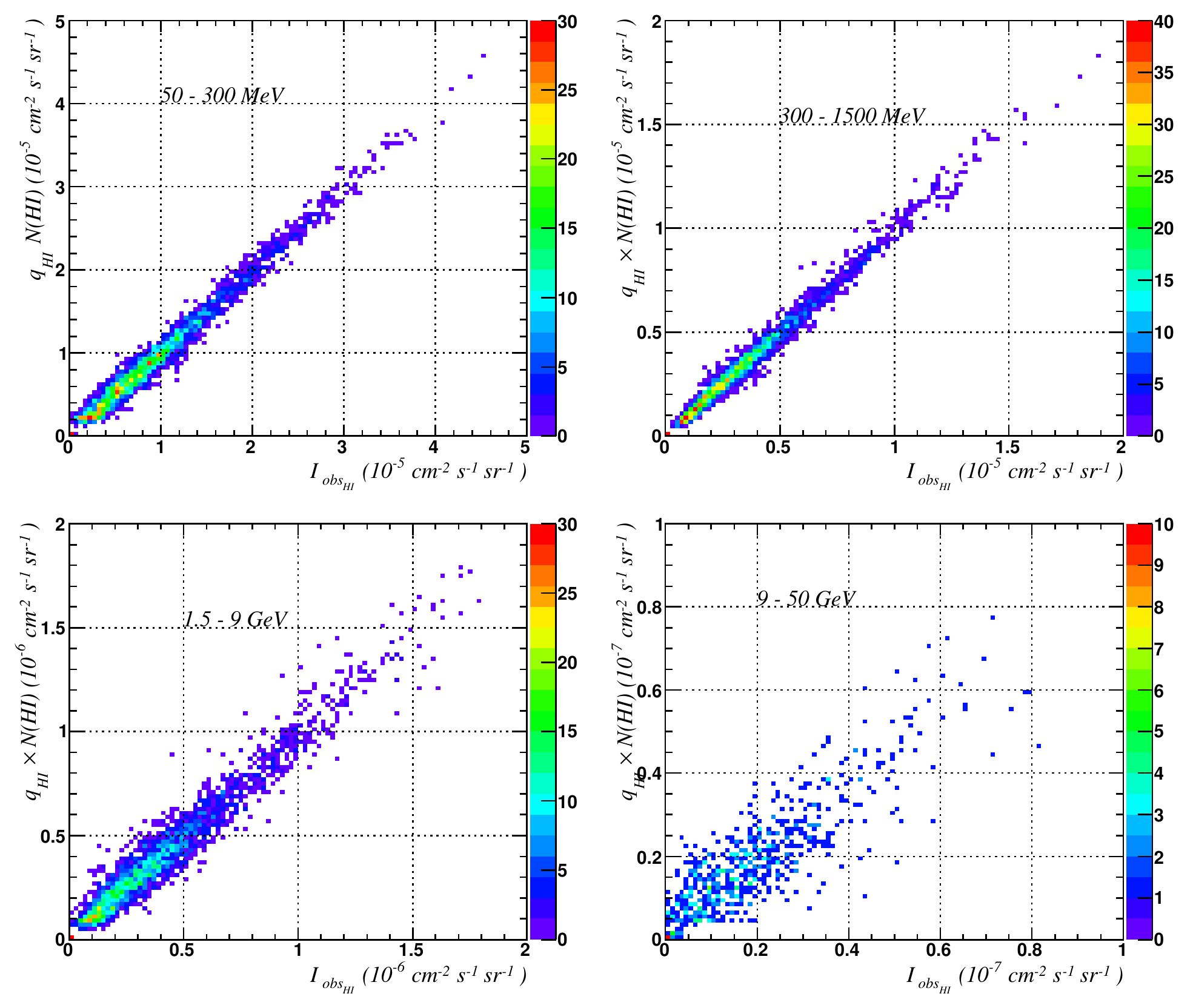}
\caption{Pixel by pixel correlation in 4 energy bands between the $\gamma$-ray intensity and H~{\sc i} column density. We multiplied the LAB survey column density by the emissivity per H~{\sc i} and plotted this value against the LAT counts not associated by the fit to any model components other than the local H~{\sc i} divided by the exposure to yield corresponding intensities. We selected pixels located within an absolute latitude range of 10$\degr$ to 70$\degr$. Each point corresponds to a pixel size of 3.7$\degr$ (7.3$\degr$ for the highest energy range).}
\label{HI_LAT_correlation_I_4plots}
\end{center}
\end{figure}

In each energy bin we calculated the map of residual LAT counts without including the local H~{\sc i} template in the model: $N_{obs_{HI}}=N_{obs}-(N_{pred} - q_{HI} \widetilde{N}$(H~{\sc i})), and divided by the exposure map and pixel solid-angle to obtain the intensity $I_{obs_{HI}}$. This intensity represents the intensity of $\gamma$ rays measured by the LAT not associated by the fit to any model components other than the local H~{\sc i}. In Figure \ref{HI_LAT_correlation_I_4plots} we show the pixel-by-pixel correlation between $I_{obs_{HI}}$ and the predicted intensity $q_{HI}N$(H~{\sc i}). In this figure we have divided our total energy range in four intervals, and averaged the intensity within 256 neighboring pixels (1024 pixels for the highest energy range) to lower the Poisson fluctuation. We observe a very good linear correlation between $\gamma$ rays and atomic hydrogen column density.

The instrument energy dispersion introduces a bias between the number of reconstructed counts in a given energy range and the number of incoming photons of corresponding energies. Above 200~MeV, for LAT event selection P7REP Clean, the combination of the incoming photon spectrum and the energy dependence of the LAT acceptance reduces the effect of energy dispersion; counts that are dispersed by an amount larger than half an energy bin width are equally distributed in bins with energies higher and lower than that of their corresponding photons. Below 200~MeV the sharp fall-off of the $\pi^{0}$ decay spectrum combined with the decrease of the effective area of the LAT breaks this equilibrium and results in an increase of counts in the lowest energy bins. We used the LAT Science Tool\footnote{Available from \url{http://fermi.gsfc.nasa.gov/ssc/}.} {\it gtrspgen} to estimate this effect for photon energies from 50~MeV to 50~GeV and observed that a correction term for the bias introduced by the energy dispersion follows the simple formula $1+ 7500E^{-2.43}$. It leads to an increase of counts of 38\% in our lowest energy range spanning from 50 to 64~MeV and of 21\% for the one from 64 to 82~MeV. The effect of the energy dispersion is less than 2\% for the range spanning from 172 to 220~MeV. We proceeded iteratively to account for the dispersion, we first used a predicted diffuse spectrum to calculate an approximate energy dispersion correction. We then applied this correction to derive the H~{\sc i} emissivity, and used the measured spectrum for a more accurate dispersion correction. We verified that no extra iteration was necessary. 

\begin{figure}
\begin{center}
\includegraphics[width=14cm]{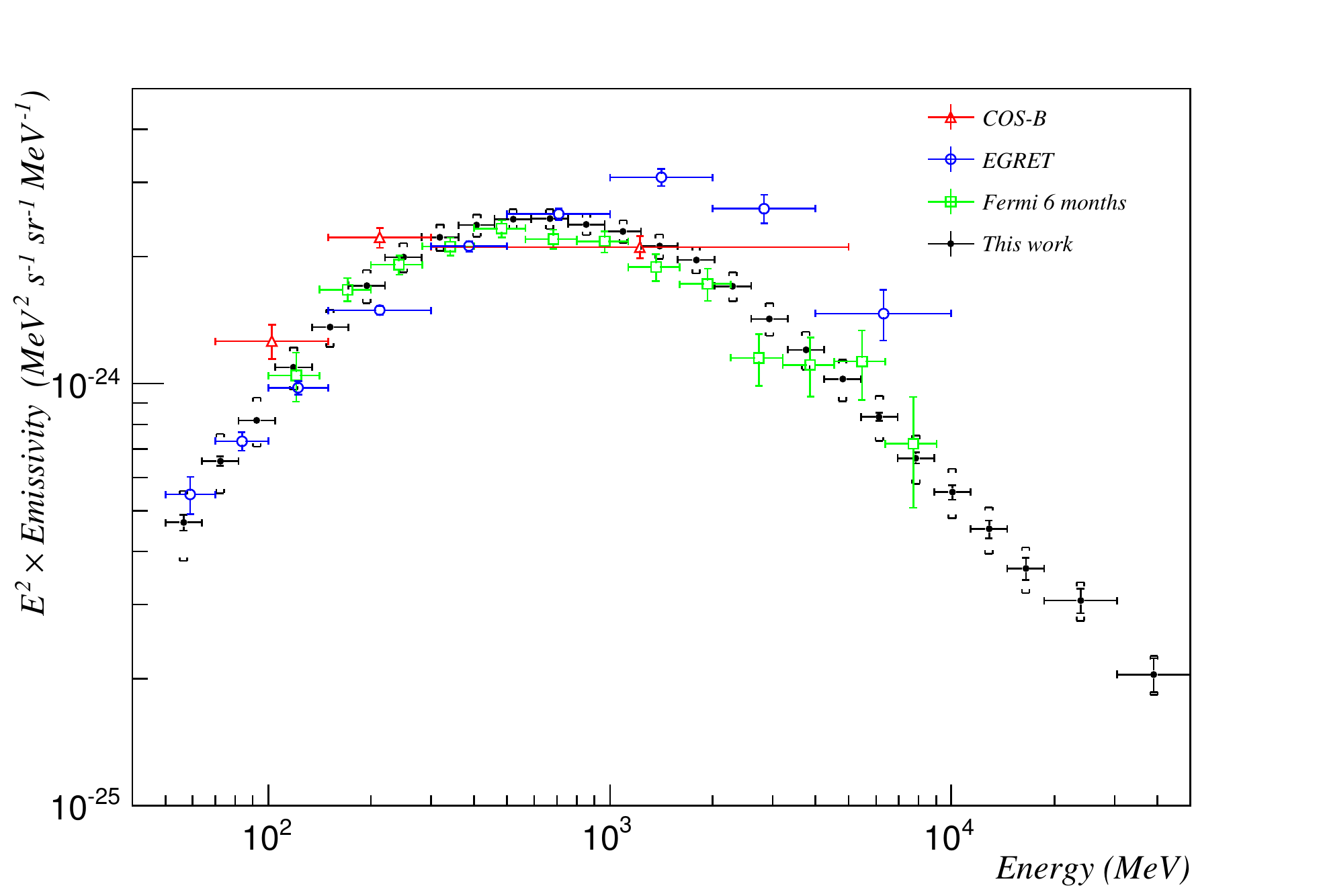}
\caption{Differential local $\gamma$-ray emissivity per H~{\sc i} versus $\gamma$-ray energy. The vertical error bars correspond to statistical uncertainties while systematics are represented by square brackets, the horizontal bars correspond to the bins energy width. The differential emissivity is compared to previous works from COS-B \citep{Strong:1988p3093}, EGRET \citep{Strong:1996p1032}, and to early {\it Fermi}-LAT results \citep{Abdo:2009p3049}.}
\label{emissivity_history}
\end{center}
\end{figure}

Taking into account this correction for the energy resolution we plotted in Figure \ref{emissivity_history} the local $\gamma$-ray differential emissivity per H~{\sc i}, obtained by dividing the emissivities $q_{HI}$ by the energy band width. We represented the statistical uncertainties (see Section 8) with vertical error bars and instrumental systematic uncertainties with brackets. We positioned the points at energies corresponding to the logarithmic mid-points of the bands and scaled the emissivities by the square of those energies. We include the differential emissivities from previous work from COS-B, EGRET, and previous {\it Fermi}-LAT results. The agreement is good except for the EGRET emissivities above 1~GeV as studied by \cite{Abdo:2009p4020}. The differential emissivities multiplied by the square of the band mid-point energy are provided in Table \ref{tbl:emissivity}.

\begin{deluxetable}{lcccccc}
\tabletypesize{\scriptsize}
\tablecaption{Local H~{\sc i} differential emissivity \label{tbl:emissivity}}
\tablewidth{0pt}
\tablehead{
\colhead{Energy band}
&\colhead{$E^2 (q_{HI}\pm stat \pm sys)$}  
&\colhead{$E^2 (q_{CO}\pm stat)$} 
&\colhead{$E^2 (q_{DNM}\pm stat)$} 
&\colhead{$E^2 (q_{H^{+}}\pm stat)$} 
&\colhead{$N_{IC_{p}}\pm stat$}
&\colhead{$E^2 (N_{iso}\pm stat)$}
\\ 
\colhead{MeV}
&\colhead{(a)}
&\colhead{(b)}
&\colhead{(c)}
&\colhead{(d)}
&\colhead{}
&\colhead{(e)}
}
\startdata
50--64   & 0.47 $\pm$ 0.02 $\pm$ 0.09 & 1.27 $\pm$0.10  & 0.00 $\pm$0.01  & 0.0 $\pm$0.6  & 0.60 $\pm$0.01  & 2.03 $\pm$0.02  \\
64--82   & 0.66 $\pm$ 0.02 $\pm$ 0.11 & 1.23 $\pm$0.09  & 0.06 $\pm$0.01  & 0.0 $\pm$0.5  & 0.71 $\pm$0.01  & 1.85 $\pm$0.01  \\
82--105   & 0.82 $\pm$ 0.02 $\pm$ 0.11 & 1.50 $\pm$0.09  & 0.25 $\pm$0.02  & 0.0 $\pm$0.4  & 0.84 $\pm$0.01  & 1.73 $\pm$0.01  \\
105--134   & 1.09 $\pm$ 0.01 $\pm$ 0.13 & 1.85 $\pm$0.08  & 0.36 $\pm$0.02  & 0.8 $\pm$0.3  & 0.90 $\pm$0.01  & 1.61 $\pm$0.01  \\
134--172   & 1.36 $\pm$ 0.01 $\pm$ 0.14 & 2.45 $\pm$0.08  & 0.49 $\pm$0.02  & 1.0 $\pm$0.4  & 1.04 $\pm$0.02  & 1.59 $\pm$0.01  \\
172--220   & 1.71 $\pm$ 0.01 $\pm$ 0.15 & 2.84 $\pm$0.07  & 0.66 $\pm$0.02  & 1.9 $\pm$0.4  & 1.18 $\pm$0.02  & 1.54 $\pm$0.01  \\
220--281   & 1.99 $\pm$ 0.01 $\pm$ 0.16 & 3.44 $\pm$0.07  & 0.75 $\pm$0.02  & 2.9 $\pm$0.3  & 1.29 $\pm$0.02  & 1.44 $\pm$0.01  \\
281--360   & 2.22 $\pm$ 0.01 $\pm$ 0.16 & 3.98 $\pm$0.07  & 0.90 $\pm$0.02  & 2.9 $\pm$0.3  & 1.43 $\pm$0.02  & 1.32 $\pm$0.01  \\
360--461   & 2.37 $\pm$ 0.01 $\pm$ 0.15 & 4.28 $\pm$0.07  & 0.96 $\pm$0.02  & 2.4 $\pm$0.3  & 1.60 $\pm$0.02  & 1.14 $\pm$0.01  \\
461--589   & 2.45 $\pm$ 0.01 $\pm$ 0.13 & 4.32 $\pm$0.07  & 1.03 $\pm$0.02  & 3.6 $\pm$0.3  & 1.72 $\pm$0.02  & 1.01 $\pm$0.01  \\
589--754   & 2.46 $\pm$ 0.01 $\pm$ 0.13 & 4.39 $\pm$0.07  & 1.03 $\pm$0.02  & 3.1 $\pm$0.3  & 1.79 $\pm$0.03  & 0.90 $\pm$0.01  \\
754--965   & 2.39 $\pm$ 0.01 $\pm$ 0.14 & 4.33 $\pm$0.07  & 1.05 $\pm$0.02  & 3.7 $\pm$0.3  & 1.90 $\pm$0.03  & 0.75 $\pm$0.01  \\
965--1235   & 2.29 $\pm$ 0.02 $\pm$ 0.14 & 4.20 $\pm$0.08  & 0.96 $\pm$0.02  & 2.8 $\pm$0.3  & 1.95 $\pm$0.04  & 0.64 $\pm$0.01  \\
1235--1581   & 2.12 $\pm$ 0.02 $\pm$ 0.14 & 3.98 $\pm$0.08  & 0.87 $\pm$0.02  & 2.9 $\pm$0.4  & 2.10 $\pm$0.04  & 0.55 $\pm$0.01  \\
1581--2024   & 1.97 $\pm$ 0.02 $\pm$ 0.14 & 3.65 $\pm$0.08  & 0.82 $\pm$0.02  & 2.1 $\pm$0.4  & 2.14 $\pm$0.05  & 0.52 $\pm$0.01  \\
2024--2590   & 1.70 $\pm$ 0.02 $\pm$ 0.13 & 3.03 $\pm$0.08  & 0.73 $\pm$0.02  & 2.2 $\pm$0.4  & 2.18 $\pm$0.06  & 0.45 $\pm$0.01  \\
2590--3314   & 1.42 $\pm$ 0.02 $\pm$ 0.13 & 2.67 $\pm$0.08  & 0.61 $\pm$0.02  & 2.0 $\pm$0.4  & 2.12 $\pm$0.05  & 0.41 $\pm$0.01  \\
3314--4242   & 1.20 $\pm$ 0.02 $\pm$ 0.12 & 2.22 $\pm$0.08  & 0.54 $\pm$0.02  & 1.2 $\pm$0.4  & 2.26 $\pm$0.06  & 0.36 $\pm$0.01  \\
4242--5429   & 1.02 $\pm$ 0.02 $\pm$ 0.11 & 1.85 $\pm$0.09  & 0.43 $\pm$0.02  & 1.6 $\pm$0.4  & 2.25 $\pm$0.06  & 0.34 $\pm$0.01  \\
5429--6947   & 0.83 $\pm$ 0.02 $\pm$ 0.10 & 1.64 $\pm$0.09  & 0.35 $\pm$0.02  & 0.5 $\pm$0.4  & 2.16 $\pm$0.07  & 0.36 $\pm$0.01  \\
6947--8891   & 0.67 $\pm$ 0.02 $\pm$ 0.09 & 1.26 $\pm$0.09  & 0.31 $\pm$0.02  & 1.8 $\pm$0.5  & 2.06 $\pm$0.08  & 0.34 $\pm$0.01  \\
8891--11379   & 0.55 $\pm$ 0.02 $\pm$ 0.07 & 1.03 $\pm$0.09  & 0.22 $\pm$0.02  & 0.1 $\pm$0.4  & 2.04 $\pm$0.09  & 0.32 $\pm$0.01  \\
11379--14563   & 0.45 $\pm$ 0.02 $\pm$ 0.06 & 0.87 $\pm$0.10  & 0.17 $\pm$0.02  & 0.7 $\pm$0.5  & 1.68 $\pm$0.15  & 0.34 $\pm$0.01  \\
14563--18638   & 0.36 $\pm$ 0.02 $\pm$ 0.05 & 0.86 $\pm$0.10  & 0.12 $\pm$0.03  & 0.3 $\pm$0.5  & 2.24 $\pm$0.15  & 0.26 $\pm$0.01  \\
18638--30527   & 0.31 $\pm$ 0.02 $\pm$ 0.03 & 0.56 $\pm$0.08  & 0.12 $\pm$0.02  & 0.8 $\pm$0.4  & 1.83 $\pm$0.17  & 0.23 $\pm$0.01  \\
30527--50000   & 0.20 $\pm$ 0.02 $\pm$ 0.02 & 0.45 $\pm$0.09  & 0.07 $\pm$0.02  & 0.6 $\pm$0.5  & 2.28 $\pm$0.16  & 0.16 $\pm$0.01  \\
\enddata         
\tablecomments{(a):10$^{-24}$ MeV$^{2}$ s$^{-1}$ sr$^{-1}$ MeV$^{-1}$ , (b):10$^{-4}$ MeV$^{2}$ s$^{-1}$ cm$^{-2}$ sr$^{-1}$ MeV$^{-1}$ (K km s$^{-1}$)$^{-1}$, (c):10$^{-2}$ MeV$^{2}$ s$^{-1}$ cm$^{-2}$ sr$^{-1}$ MeV$^{-1}$ mag$^{-1}$, (d): 10$^{-24}$ MeV$^{2}$ s$^{-1}$ sr$^{-1}$ MeV$^{-1}$, (e):10$^{-3}$ MeV$^{2}$ s$^{-1}$ cm$^{-2}$ sr$^{-1}$ MeV$^{-1}$ }
\end{deluxetable}

\begin{figure}
\begin{center}
\plotone{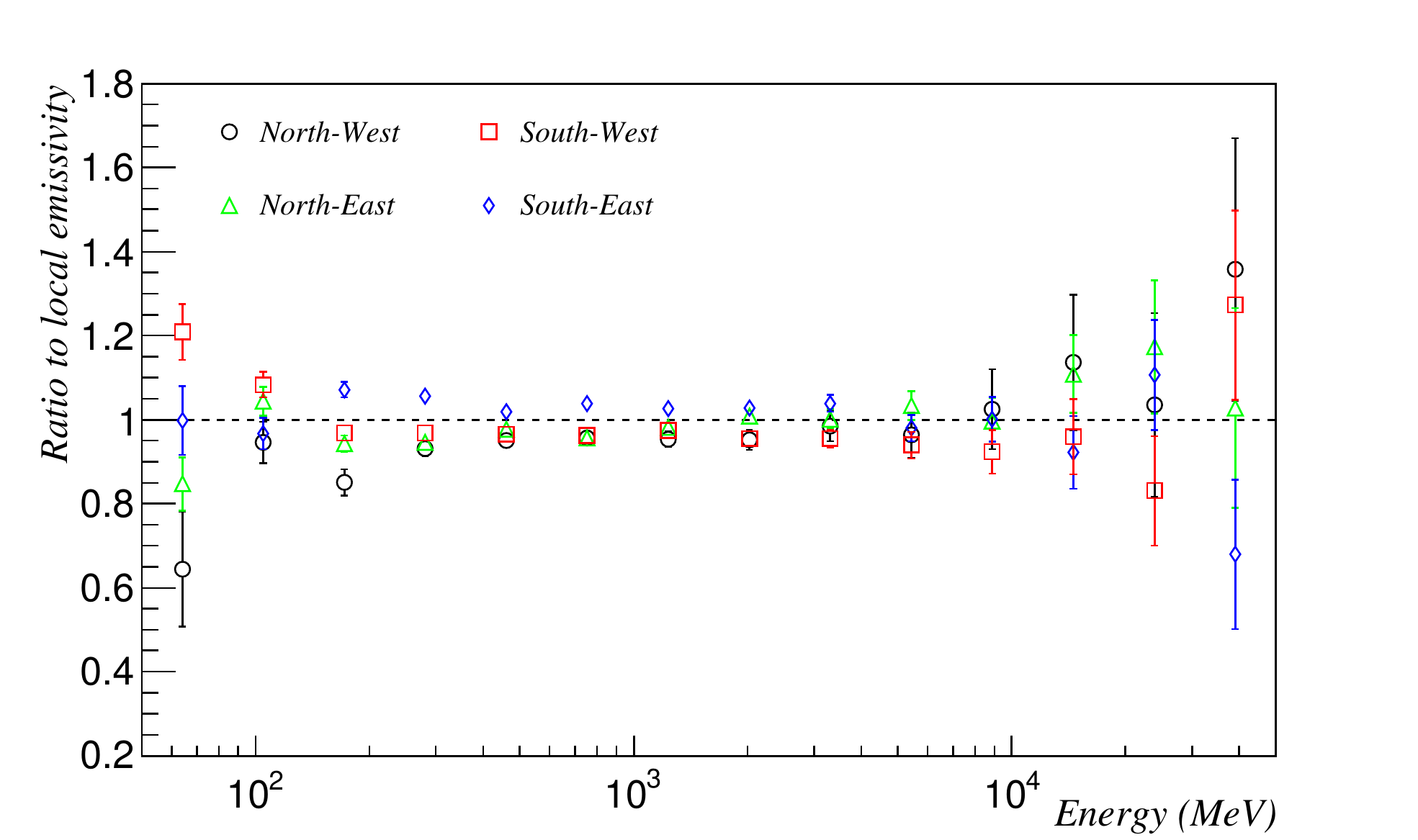}
  \caption{Ratio of the emissivity in four sky quadrants to the emissivity in the whole region analyzed versus $\gamma$-ray energy. Only statistical uncertainties are shown.  See the text for the definitions of the quadrants.}
\label{emiss_4_quadrants}
\end{center}
\end{figure}

We investigated possible variation of the emissivity spectrum across the sky. We divided our region of interest in four quadrants and restricted the fit of Equation \ref{eqRing} to the quadrant North-West: $10\degr \leq b \leq 70\degr$ and $180\degr \leq l \leq 360\degr$, South-West: $-70\degr \leq b \leq -10\degr$  and $180\degr \leq l \leq 360\degr$ North-East: $10\degr \leq b \leq 70\degr$ and $0\degr \leq l \leq 180\degr$, and South-East: $-70\degr \leq b \leq -10\degr$ and $0\degr \leq l \leq 180\degr$. We show in Figure \ref{emiss_4_quadrants} the emissivities in the four quadrants divided by the emissivity obtained previously in the whole region. We obtain a good agreement between the emissivities in the quadrants, indicating no significant local CR asymmetry across the sky.

\begin{figure}
\begin{center}
\includegraphics[width=14cm]{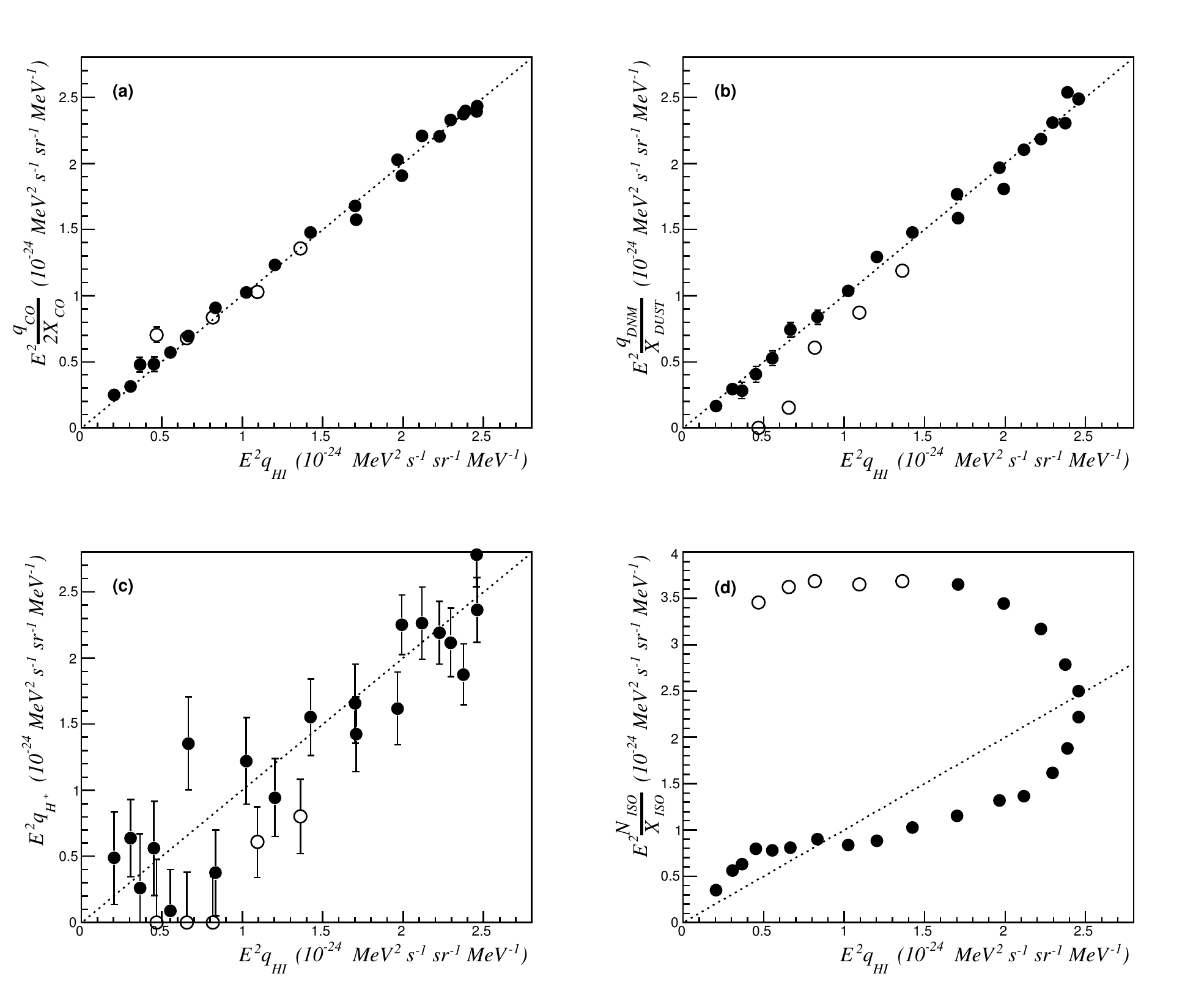}
\caption{Comparison between $q_{HI}$ and (a) $q_{CO}$, (b) $q_{DNM}$, (c) $q_{H^{+}}$, and (d) the isotropic normalization. The emissivities are scaled by the square of the energy at which the emissivity was measured. The open circle symbol represents the emissivities derived for energies below 200~MeV. The emissivities measured in the molecular, ionized and DNM are scaled to the emissivity per H~{\sc i} so that identical emissivity spectra in those different phases would be aligned on the dashed median line. We observe a similar spectral shape for the hydrogen emissivity in the four components: atomic, molecular, ionized, and in the DNM. As expected we do not observe any correlation between the isotropic normalization factor and the emissivity since no correlation is expected between extragalactic $\gamma$ rays and those issued from the atomic hydrogen. Error bars correspond to statistical errors only.}
\label{fig_XCO}
\end{center}
\end{figure}

We verified that the $\gamma$-ray emissivities measured in the local atomic hydrogen are compatible with the ones extracted from the molecular hydrogen, DNM gas and ionized hydrogen. Similarly to what we did for H~{\sc i} we obtained from the fit of Equation \ref{eqRing} the emissivities associated to $W$(CO) ($q_{CO}$) and to the DNM template ($q_{DNM}$) and derived the molecular hydrogen-to-CO conversion factor $X_{CO}=q_{CO}/2q_{HI}$ and the dust-to-gas ratio $X_{DUST}=q_{DNM}/q_{HI}$. Restricted to the energy range above 200~MeV, where the correlation between the CO template and point sources is weaker, we obtained $X_{CO}=(0.902\pm0.007) \times 10^{20}$ cm$^{-2}$ (K km s$^{-1}$)$^{-1}$ and $X_{DUST}=(41.4\pm0.3) \times 10^{20}$ cm$^{-2}$ mag$^{-1}$. The values quoted here correspond to clouds and dust within the absolute latitude range 10$\degr$ to 70$\degr$. The uncertainties do not account for systematics (see Section ~\ref{sec:systematic_errors}). 
Figure~\ref{fig_XCO} shows the correlation between the $\gamma$-ray emissivity in the atomic and the molecular phase (a) and DNM (b) obtained from $q_{CO}$ and $q_{DNM}$ scaled with the ratios $X_{CO}$ and $X_{DUST}$. As in Figure \ref{emissivity_history} we scaled the emissivities by the square of the energy. We observe a good linear correlation (Figure~\ref{fig_XCO}a-b) indicating that the spectral shape of the hydrogen emissivity is similar when the hydrogen is atomic, a molecule and in the gas that is traced solely by dust. To verify the method, we proceeded similarly with the isotropic template by scaling $N_{iso}$ by a factor $X_{iso}$ to match the range of the emissivity per H~{\sc i}.  As expected there is no linear correlation between the isotropic normalization and the atomic hydrogen emissivity (Figure~\ref{fig_XCO}d).

In Figure \ref{fig_XCO}c we show for the first time the linear correlation between $\gamma$-ray emissivities measured in H~{\sc i} and in H$^{+}$ ($q_{H^{+}}$). As explained in Section \ref{sec:Ionized_hydrogen} we converted the ionized hydrogen free-free emission measure to H$^{+}$ column density.  This conversion depends on the the electron effective density parameter $n_{eff}$ that we adjusted to match the emissivity in H$^{+}$ to the one in H~{\sc i}. We obtained the best fit for $n_{eff}$=4.6~cm$^{-3}$ which is in the expected range of expected \citep{Sodroski:1989p4012}. That corresponds to $N$(H$^{+}$)$=1.3\times10^{20} I_{ff}$ where $I_{ff}$ is expressed in mK and $N$(H$^{+}$) in atoms cm$^{-2}$ and to an average ratio $N$(H$^{+}$)/$N$(H~{\sc i}) equal to 0.015. This proportion of H$^{+}$ to H~{\sc i} atoms is notably less than that estimated from dispersion measures (see Section \ref{sec:Ionized_hydrogen}). The fit may be biased by the strongly emitting H~{\sc ii} regions that are incorrectly reconstructed by our average electron temperature. The remaining $\gamma$-ray emission correlated with H$^{+}$ might then be associated to other smoothly varying templates of Equation \ref{eqRing} like the IC or the isotropic distribution. It is also possible that the H$^{+}$ is traced together with the DNM by the dust or that it follows the H~{\sc i} distribution. In this case the $\gamma$-ray emissivity per H~{\sc i} measured in this work would be overestimated by the amount of H$^{+}$ mixed in H~{\sc i}. The large uncertainties at low energies of the hydrogen emissivities measured in H$^{+}$ prevented the study of screening effect in the Bremsstrahlung production function \citep{Hunter:1997p329}.

\section{Systematic Errors}
\label{sec:systematic_errors}
We investigated three major sources of systematic errors: the H~{\sc i} spin temperature ($T_{S}$), the modeling of the IC, and the absolute determination of the LAT effective area.

When deriving $N$(H~{\sc i}) from the LAB radio survey we corrected for the 21-cm H~{\sc i} line opacity which depends on its $T_{S}$. Since CRs interact with H~{\sc i} independently of $T_{S}$ and since the Galaxy is optically thin to $\gamma$ rays at energies considered here we investigated the uncertainty related to $T_{S}$ by comparing the LAT counts map with models based on $N$(H~{\sc i}) derived with various $T_{S}$ values. The thinness of the local H~{\sc i} does not make $N$(H~{\sc i}) sensitive enough to variation of $T_{S}$ therefore we selected a region toward the Galactic anticenter ($90\degr \le l \le 270\degr$, $\left| b \right| <70\degr$) to obtain the H~{\sc i} $T_{S}$ that gives the best fit to the data. We investigated seven values of $T_{S}$ from 90~K to 400~K. For each temperature, we derived an H~{\sc i} column density and a DNM map. We used those maps in a fit to the {\it Fermi}-LAT counts in the anticenter region. We observed that the likelihood was maximum for the model derived with  $T_{S}$=140~K. We adopted this temperature to derive $N$(H~{\sc i}) for the whole Galaxy. We used the values 125~K and 170~K to estimate the uncertainty in the emissivity related to the $T_{S}$. This uncertainty range is larger than the one obtained by varying the maximum likelihood by 1 i.e., the statistical uncertainty in the fit parameter. This range corresponds to values of $T_{S}$ above which a noticeable discrepancy is observed in the fit residual map. We used in Equation \ref{eqRing} $N$(H~{\sc i}) derived with a $T_{S}$ of 125, 140 and 170~K and compared in each energy bin the resulting emissivities. We obtained a systematic uncertainty related to Ts of 2\% corresponding to the average variation of those emissivities. Studies of H~{\sc i} absorption against background radio sources have shown that $T_{S}$ is not uniform in the multi-phase ISM \citep{Heiles:2003p1829,Kanekar:2011p4066}, but we assumed here a uniform $T_{S}$ and corrected this approximation by representing in the model maps from positive and negative residuals of the total dust column density.

As shown by \cite{Ackermann:2012p2978} there is no unique set of GALPROP propagation parameters that can model the Galactic diffuse emission in the {\it Fermi}-LAT sky. Since we rely on GALPROP for the morphology of the IC map, we investigated the effect of three sets of GALPROP parameters labelled 54\_z10G4c5rS, 54\_77Xvarh7S and $^SY^Z6^R30^T150^C2$ \citep{Ackermann:2012p2978}. This method does not account for all the uncertainties related to the Galprop IC map but since the spatial structures of $N$(H~{\sc i}) allow for a good separation from the smooth morphology of the IC map, we do not need to input into our fit an accurate IC template. As an illustration, when we do not include into the fit any IC template, the HI emissivity varies by only 20\%. The effect of using the three different IC models is limited, from 3\% at low energy to 1\% at higher energy. We used recent IC distribution predictions obtained with the version $^SY^Z6^R30^T150^C2$ for the final $\gamma$-ray fit and added in quadrature to the systematic uncertainties the variation obtained from the other parameter sets.

Finally we used the following values for the systematic uncertainties related to LAT effective area: 20\% at 50~MeV, 10\% at 100~MeV, 5\% at 560~MeV, 10\% at 10~GeV and we extrapolated linearly those values using the logarithm of the energy. Those values are more conservative than the recommended\footnote{\url{http://fermi.gsfc.nasa.gov/ssc/data/analysis/LAT\_caveats.html}} ones. The LAT effective area uncertainties dominate the uncertainties related to $T_{S}$ and IC in the whole energy range. 

There are as many $\gamma$ rays originating from interstellar emission within 6$\degr$ of the Galactic plane as of the rest of the Galaxy. At low energy, where the PSF is broad, counts corresponding to $\gamma$ rays originating from the plane spread out to higher latitude. We incorporated into Equation \ref{eqRing} the inner gas annuli that model the Galactic plane to account for this spreading. We tested the stability of the fit with four cuts corresponding to absolute latitudes higher than 5$\degr$, 7$\degr$, 10$\degr$ and 15$\degr$ and found no statistically significant variation. We also excluded all $\gamma$ rays detected at absolute latitudes higher than 70$\degr$, where the number of counts coming from collisions with H~{\sc i} is four times less than those of extragalactic and instrumental origins. For the final emissivity fit we applied a cut at 10$\degr$ which allows a good template separation by the fit while excluding most of the Galactic plane.

Incorrect modeling of large scale structures like Loop~{\sc i} or the {\it Fermi} Bubbles would bias the fit. To investigate the influence of those large structures, we restricted the fit to a region away from them at longitudes between 50$\degr$ and 280$\degr$. We observed no noticeable difference for the local H~{\sc i} emissivity. This test, in a region where the intensity of $\gamma$ rays from IC is much lower than from $\pi^{0}$ decay, also indicates that correlation between $I_{IC_{p}}$ and $N$(H~{\sc i}) maps have negligible effects on the measured emissivities.

\section{Interpretation}
\subsection{Production cross-section}

Most $\gamma$ rays with energies between 100~MeV and 50~GeV originate from the decay of $\pi^{0}$ produced in hadronic collisions when CR protons with energies from 0.5 to 10$^3$~GeV interact with ISM nuclei. At those energies, the vast majority of $\pi^{0}$ are produced with a low transverse momentum transfer through soft interactions where the large strong force coupling constant prevents the use of a perturbative quantum chromodynamics approach. This regime is poorly understood and the $\pi^{0}$ production cross-section is obtained mainly through phenomenological models. In this work we have tested three $\gamma$-ray production cross-sections:
 
In \cite{Kamae:2005p2602} and \cite{Kamae:2006p2590} the authors have used the accumulated knowledge of high-energy accelerator physics to interpret the $\pi^{0}$ production cross-section for astrophysical purpose. They used the event generator Pythia \citep{Sjostrand:2006p3632} to evaluate the non-diffractive component above a p-p center of mass energy of 10~GeV where phenomenological fits based on the Regge theory \citep{Donnachie:1992p3628} are believed to be valid. At lower energy they used the scaling formula of \cite{Stephens:1981p3627} with parameters based on \cite{Blattnig:2000p3630} but readjusted to account for the additional model components. Pythia offers an implementation of phenomenological QCD that effectively takes care of higher order effects (e.g., violation of Koba-Nielsen-Olesen scaling) that covers most of the deviation from a simple scaling law in the non-diffractive cross-section. Single and double diffraction were considered as well as resonances at 1232~MeV/$c^2$ and around 1600~MeV/$c^2$. This model provides a good fit of the experimental $\pi^{0}$ inclusive cross-section.

A similar approach was used by \cite{Huang:2007p2906} where the cross-section for proton energies above 20~GeV was calculated using the Monte Carlo particle collisions code DPMJET-III \citep{Roesler:2001p4021} that describes the nuclear interactions with the two-component Dual Parton Model again based for soft processes on the Regge phenomenology. At lower proton energies \cite{Huang:2007p2906} used the same parametric model as \cite{Kamae:2006p2590} and incorporated to their model resonances at 1232~MeV $c^{-2}$ and 1600~MeV $c^{-2}$. The contribution of  $\eta$ mesons (classified with the direct products in \cite{Huang:2007p2906}) appears to be overestimated by DPMJET-III; we reduced it by a factor 8 to match the experimental cross-section measured in \cite{AguilarBenitez:1991p4004}.

The approach chosen by \cite{Shibata:2014p4188} is different and based only on $\gamma$-ray production cross-sections obtained in p-p collisions spanning the range of energies from the threshold pion production to energies of the Large Hadron Collider experiments. The cross-section is parametrized with an empirical formula depending on the $\gamma$-ray energy and transverse momentum as well as four physical parameters that are adjusted so that the empirical cross-section and deduced quantities fit the experimental data \citep{Shibata:2013p3959,Sato:2012p3926,Shibata:2007p3958,Suzuki:2005p3957}.

{\it Fermi} detects $\gamma$ rays resulting not only from p-p interactions but also from the interactions of CR nuclei with the ISM nuclei. We used \cite{Mori:2009p341} to scale $\sigma_{pp}$ to nucleus-nucleus cross-section. Table \ref{tbl:mori_scale} gives the scaling factors of \cite{Mori:2009p341} obtained by comparing the $\gamma$-ray yield from p-p, p-nucleus, nucleus-p and nucleus-nucleus collisions calculated with DPMJET-III. We used the values calculated at 10~GeV/nucleon and neglected their dependence on energy which is of the order of a few percent \citep{Kachelriess:2014p4196}. Table \ref{tbl:mori_scale_ISM} lists the scaling factor for collisions induced by proton and Helium CRs after including an ISM relative abundance of $p:He:CNO:MgSi:Fe = 1:0.096:1.38\times 10^{-3}:2.11\times10^{-4}:3.25\times10^{-5}$ \citep{Meyer:1985p3637,Mori:2009p341}. From this table we see that: $\sigma_{p,ISM}=1.387 \sigma_{pp}$ and $\sigma_{He,ISM}=5.12 \sigma_{pp}$. CRs heavier than Helium colliding with the ISM also contribute to the total $\gamma$-ray yield. Their relative abundances were estimated at 10~GeV from Table 1 of \cite{Honda:2004p3922} as $p:CNO:MgSi:Fe = 1:0.0032:0.0016:0.00037$ at 10~GeV. Using these abundances together with cross-section scaling factors of Table ~\ref{tbl:mori_scale}, we obtained a correction equivalent to $0.116 \sigma_{pp}$ that we added to $\sigma_{p,ISM}$ so that $\sigma_{p+heavy,ISM}=1.503 \sigma_{pp}$. For information, using a CR ratio of 0.055 between Helium and proton CR fluxes at 10 GeV, we obtain a total enhancement of 1.78. Here we do not use this value directly since we treated protons and Helium CRs separately.

\begin{deluxetable}{lccccccc}
\tabletypesize{\scriptsize}
\tablecaption{Multiplication factors at 10~GeV/nucleon from \cite{Mori:2009p341} by which the p-p cross-section is scaled to obtain a nucleus-nucleus cross section.  \label{tbl:mori_scale}}
\tablewidth{0pt}
\tablehead{  
\multicolumn{2}{c}{Projectile}   &  \multicolumn{5}{c}{Target} \\
\cline{3-7} \\
\colhead{}&\colhead{}&\colhead{H}&\colhead{He}&\colhead{CNO}&\colhead{Mg-Si}&\colhead{Fe}  }
\startdata
H      &    &  1. & 3.81 & 11.6 & 20.1 & 38.9        \\
He     &    &  3.68 & 14.2 & 42.3 & 73.4 & 143.1     \\
CNO    &    &  11.7 & 42.5 & 120.6 & 204.2 & 386.3   \\
Mg-Si  &    &  20.3 & 73.2 & 204.2 & 343.1 & 628.8   \\
Fe     &    &  38.8 & 142. & 384.4 & 634. & 1067.    \\
\enddata
\end{deluxetable}

\begin{deluxetable}{lccccccc}
\tabletypesize{\scriptsize}
\tablecaption{Multiplication factors of Table ~\ref{tbl:mori_scale} scaled by ISM abundance. \label{tbl:mori_scale_ISM}}
\tablewidth{0pt}
\tablehead{  
\multicolumn{2}{c}{Projectile}   &  \multicolumn{6}{c}{Target} \\
\cline{3-7} \\
\colhead{}&\colhead{}&\colhead{H}&\colhead{He}&\colhead{CNO}&\colhead{Mg-Si}&\colhead{Fe}&\colhead{Sum}  }
\startdata
H      &    & 1. & 0.366 & 0.016 & 0.004 & 0.001 & 1.387 \\
He     &    &3.68 & 1.363 & 0.058 & 0.015 & 0.005 & 5.12 \\
\enddata
\end{deluxetable}

The $\gamma$ rays detected by {\it Fermi}-LAT at energies relevant for this work are also produced when CR electrons and positrons of about 0.1 to 10~GeV are bremsstrahlung scattered by ISM nuclei. We used \cite{Gould:1969p2474} to calculate this bremsstrahlung radiation. For this we used tabulated $\phi1$ and $\phi2$, and we assumed $\sigma_{bremss}(H_{2})=2\sigma_{bremss}(H~I)$.

\subsection{Comparison with Proton and Helium Heliospheric Spectra}
We first compared the $\gamma$-ray emissivity per hydrogen atom measured by {\it Fermi}-LAT with the one calculated from CR fluxes measured directly (proton and Helium) and indirectly (electron). 
To calculate the emissivities we folded the $\gamma$-ray production cross-sections with the proton, Helium and electron flux. We used the spectra of proton and Helium gathered by PAMELA between 2006 and 2008 \citep{Adriani:2011p4018} and ATIC-2 \citep{Panov:2006p4024}. Since PAMELA and AMS-01 \citep{Aguilar:2002p3620} measurements are in agreement below 10~GeV, we extended the PAMELA proton spectrum to energies lower than 0.4~GeV using the two lowest-energy points of AMS-01. 

The electrons that produce $\gamma$ rays at {\it Fermi}-LAT energies also spiral around the Galactic magnetic field lines and produce radio synchrotron with frequencies from a few MHz to approximately ten GHz. This range corresponds to the one \cite{Webber:2008p2756} used to derive an electron LIS. We folded the bremsstrahlung cross-section with the CR electron local spectra of \cite{Webber:2008p2756} globally scaled by a factor 0.64 to match the {\it Fermi}-LAT electron-positron spectrum of \cite{Ackermann:2010p4023} above 10~GeV.

\begin{figure}
\begin{center}
\plotone{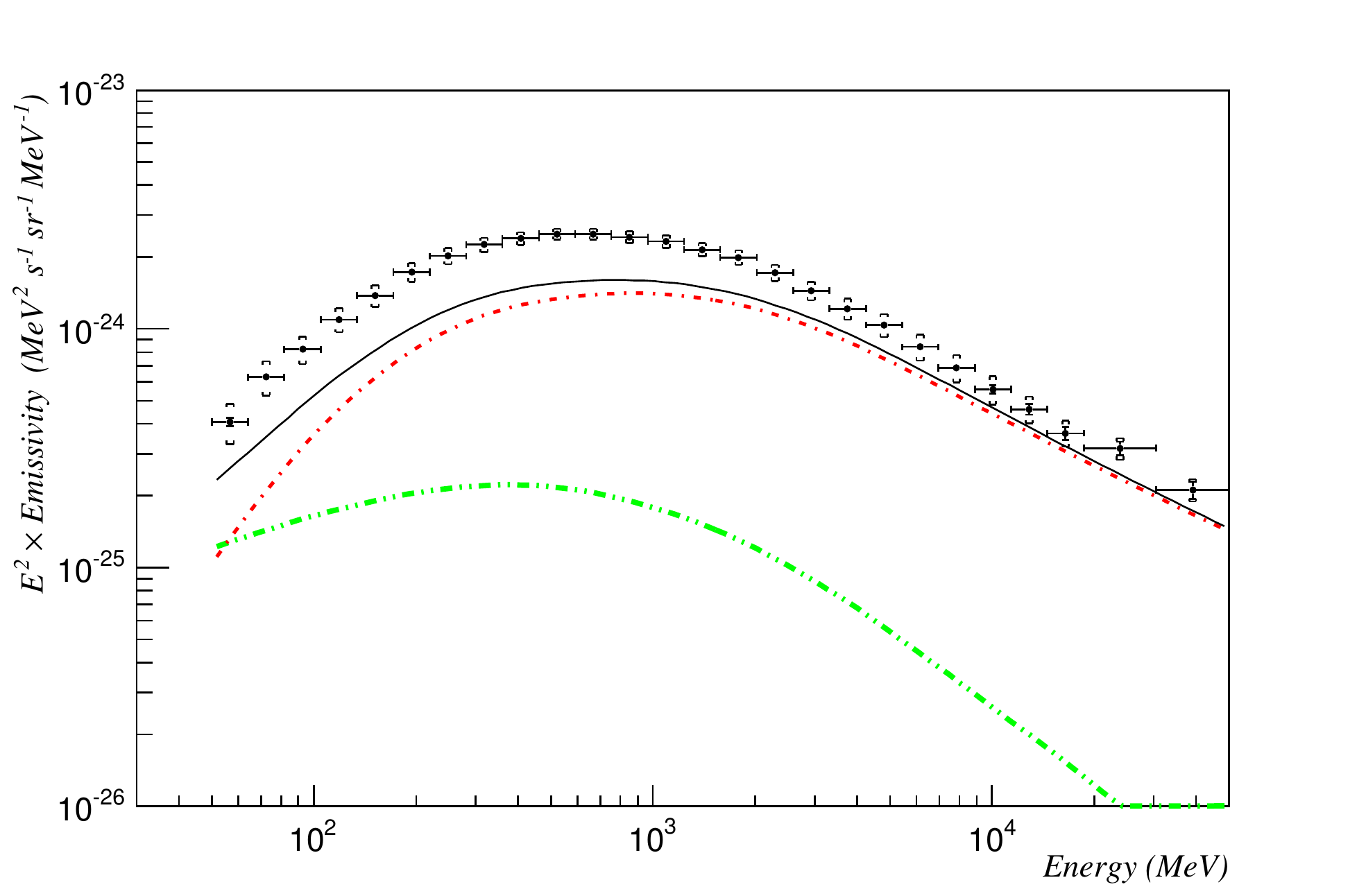}
\caption{Differential local $\gamma$-ray emissivity per H~{\sc i} versus $\gamma$-ray energy. We calculated the contribution associated with hadronic collisions by folding directly the proton and Helium spectra detected by PAMELA and ATIC-2 with the $\gamma$-ray production cross-section of Kamae {\it et al.} scaled by the multiplication factors provided in Table \ref{tbl:mori_scale_ISM} (red dash-dotted). We calculated the bremsstrahlung from the electron LIS of Webber {\it et al.} scaled by a factor 0.64 (green dash-double-dotted). The solid line represents the sum of those contributions; it does not reproduce the LAT emissivities because of the CR modulation by the Sun. No free parameters were adjusted for this graph. \label{emissivity_phi0}}
\end{center}
\end{figure}

In Figure \ref{emissivity_phi0} we show the $\gamma$-ray emissivity calculated with the Kamae {\it et al.} cross-section compared the results derived here from {\it Fermi}-LAT observations. We decomposed the total emissivity into contributions from bremsstrahlung and hadronic collisions. For this graph we did not solar-demodulate the proton or Helium spectra, but instead used directly the fluxes from PAMELA and ATIC-2. We used a power-law interpolation between the measurements in calculating the folding. We observe a disagreement between the predicted and the measured emissivities that can be attributed to the solar modulation of proton and Helium CRs. 

\begin{figure}
\begin{center}
\plotone{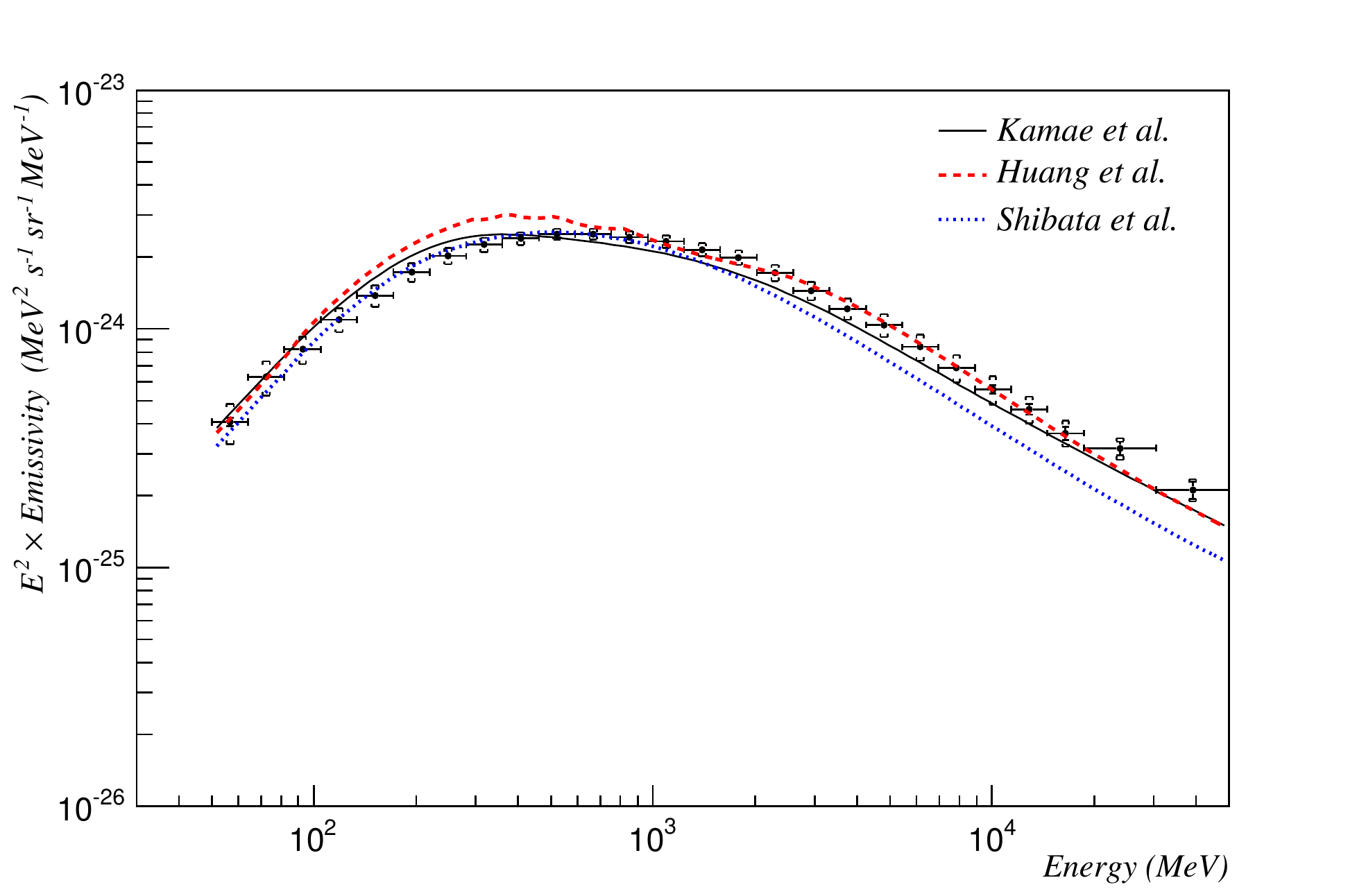}
\caption{Differential local $\gamma$-ray emissivity per H~{\sc i} compared to predictions based on \cite{Kamae:2006p2590} (black solid), \cite{Huang:2007p2906} (red dashed),
   and \cite{Shibata:2013p3959} (blue doted) $\gamma$-ray production cross-sections. Here we apply a force-field approximation to the CR proton and Helium spectra with the potential of $\Phi$ = 500~MV. \label{emissivity_cs}}
\end{center}
\end{figure}

In Figure \ref{emissivity_cs} we compare our results with the emissivities calculated using the production cross-sections of Kamae {\it et al.}, Huang {\it et al.}, and Shibata {\it et al.}. We used the force-field approximation \citep{Gleeson:1968p3685} to demodulate the proton and Helium spectra with an arbitrary  modulation potential of $\Phi$ = 500~MV. The agreement between the observed and modeled emissivity is greatly improved by taking into account the modulation due to the Sun. We observe a spread with a standard deviation of the order of 10\% between the calculated emissivities due to the differences in the production cross-sections.

\subsection{Derivation of the Proton, Helium, and Electron Local Interstellar Spectra}
We saw above that solar-demodulated heliospheric fluxes folded with the $\pi^{0}$ cross-section result in $\gamma$-ray emissivities similar to the ones we measured. Going one step further, we deduced the LISs for proton, Helium and electron-positrons by fitting variable LIS spectral shapes to direct measurements of modulated CR spectra as well as to the H~{\sc i} emissivity measurements. 

Diffusive shock acceleration theory predicts that the particle distribution function and the number density of protons follow a power law in momentum $dN/dp \propto p^{-P_2}$. Since flux and density are related by $dF/dE_{kin} = \beta c/4\pi ~ dN/dE_{kin}$ where $\beta=v/c$, and since $dp/dE_{kin} = 1/\beta$, we can write $dF/dE_{kin} =  c/4\pi ~ dN/dp \propto p^{-P_2}$ where $E_{kin}$ is the CR kinetic energy per nucleon. Therefore, as noted in \cite{Dermer:2012p3491} and \cite{Dermer:2013p4022}, if the differential flux is plotted versus the momentum, one expects to observe a power-law spectrum. For protons and Helium we adopted a spectral model similar to the one used to describe BESS fluxes \citep{Shikaze:2007p3492}: $dF/dE_{kin}=A\beta^{P_1}(p/p_{0})^{-P_2}$ where $p$ is the momentum per nucleon and $p_{0}=1~GeV/c$. The factor $\beta^{P_1}$ allows for a possible departure from a power-law spectral shape at low energy. We tested several forms; this spectral shape gave the best compromise between the number of free parameters ($A,P_1,P_2$) and goodness of the fit to the heliospheric fluxes. For electron-positrons  a similar formula did not provide a sufficient hardening at low energies since $\beta \sim 1$. Based on the measurements of AMS-01 we chose the following form that allows a harder spectrum below 1~GeV: $dF/dE_{kin} =A(E_{kin}+(E_{kin}+P_1)^{-0.5})^{-P_2}$.
With 10 free parameters (3$\times$3 for the Helium, proton and electron-positron spectral forms plus the force-field solar modulation potential) we minimized the sum of four $\chi^2$ values for four fits:  the proton spectral form to a combination of heliospheric proton fluxes from PAMELA and ATIC-2 from 0.5 to 200~GeV, the Helium spectral form to Helium fluxes also from PAMELA and ATIC-2 in the same energy range, the electron-positron spectral form to the fluxes measured by {\it Fermi}-LAT from 7 to 200~GeV, and the corresponding emissivities to our {\it Fermi}-LAT $\gamma$-ray emissivities. While the PAMELA and ATIC-2 proton and Helium energy ranges are relevant for $\pi^{0}$ production, the one of the {\it Fermi}-LAT electron-positron corresponds to $\gamma$-ray energies where the bremsstrahlung component is negligible. The fit to the LAT electron-positron measurement only fixes the asymptotic shape of the electron-positron form. We decided to not use any direct electron observation at lower energy to fit the electron-positron form to simplify the calculation of the solar modulation. The electron-positron spectrum therefore depends strongly on the H~{\sc i} emissivity as well as on the function we choose to represent its shape. For the evaluation of $\chi^2$ we included systematics and statistical uncertainties in the local emissivity, as well as the published uncertainties for PAMELA and ATIC-2 data. The CR spectral forms were folded with the $\pi^{0}$ cross-section of \cite{Kamae:2006p2590}; we propagated an error of 20\% on this cross-section to account for the spread between models that we observed in the previous section and for the energy dependence of the scaling factors that we neglected. The parameters resulting from the fit are given in Table~\ref{table_resultat_fit}. In addition, we find a solar modulation potential corresponding to the data-taking period of PAMELA considered here of (580 $\pm$ 30)~MV. This value is higher than indirect measurements of ground-based neutron detector \cite{Usoskin:2011p4190} or of the ACE/CRIS spectrometer \footnote{\url{http://www.srl.caltech.edu/ACE/ACENews/ACENews155.html}} which predict a modulation potential decreasing from 550~MV to 350~MV between July 2006 and February 2008. A possible origin for this descrepency is the under-prediction of the model above 1~GeV that the fit compensates for by increasing the modulation potential. As a consequence the experimental emissivities at low energies are over-predicted.

\begin{figure}
\begin{center}
\includegraphics[width=16cm]{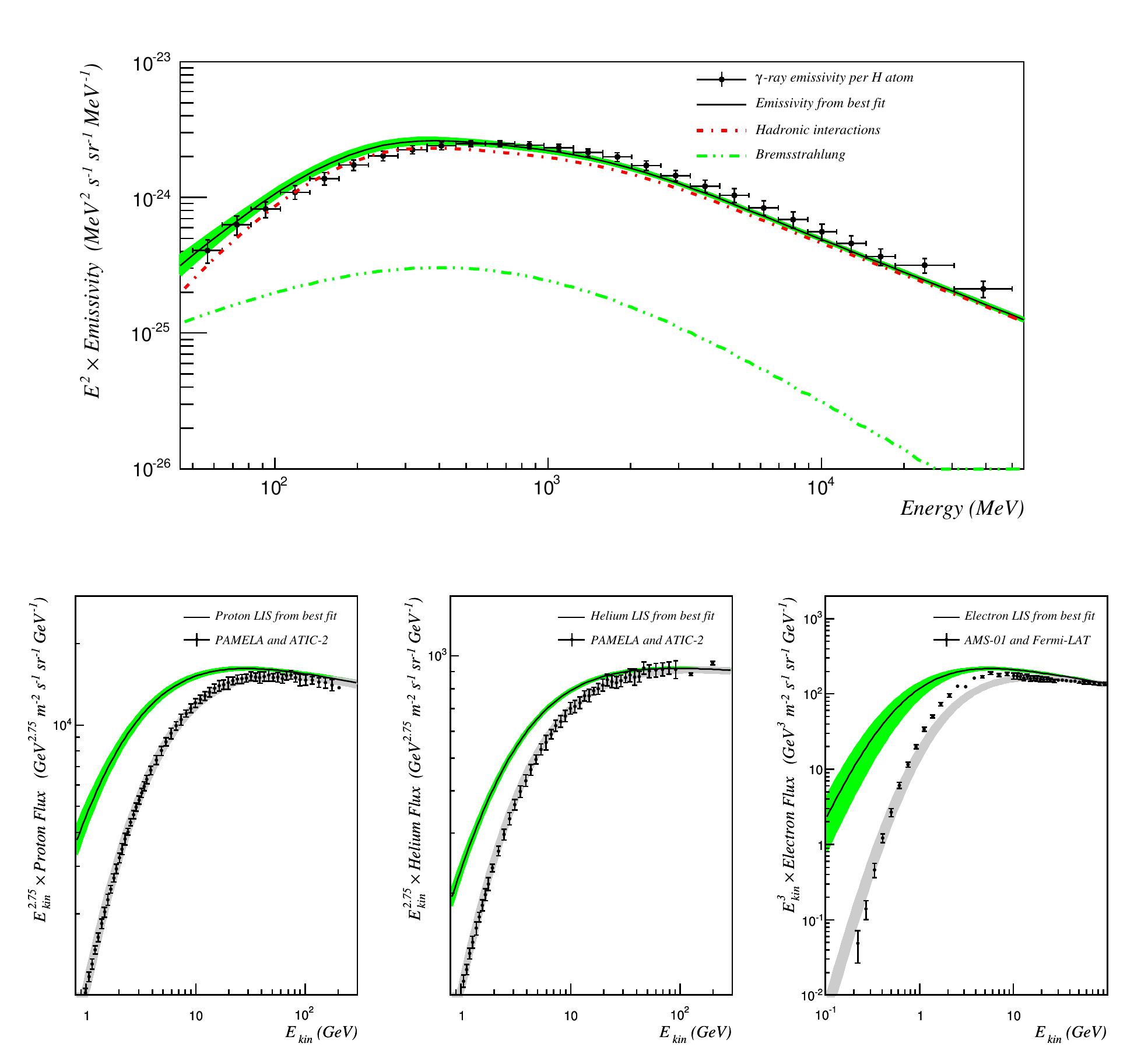}
\caption{Proton, Helium, and electron-positron LISs resulting from the fit combining the LAT emissivities and heliospheric data. Top: LAT (black points) and predicted (solid line) emissivity calculated with the \cite{Kamae:2006p2590} cross-section and proton, Helium, and electron fluxes represented by a spectral form. It is decomposed into an hadronic (red dash-dotted) and a bremsstrahlung (green dash-double-dotted) contribution. The green contour corresponds to 1$\sigma$ variations in the fit parameters. Bottom: The black line represents the fitted spectral form for proton (left), Helium (middle) and electron-positron CRs (right). The green contour corresponds to 1$\sigma$ variations in the CR fit parameters. When we apply the solar modulation to this contour we obtain the grey contour. Heliospheric fluxes for proton and Helium are provided by PAMELA \citep{Adriani:2011p4018} and ATIC-2 \citep{Panov:2006p4024} (above 100~GeV) observations. In the lower right plot corresponding to the electron flux we show data from AMS-01 \citep{Aguilar:2002p3620} together with {\it Fermi}-LAT electron-positron fluxes (above 10~GeV) but only the latter were used for the fit. We scaled the proton and Helium fluxes by $E_{kin}^{2.75}$ and the electron-positron fluxes by $E_{kin}^3$.}
\label{emissivity_contour}
\end{center}
\end{figure}

\begin{figure}
\begin{center}
\includegraphics[width=17cm]{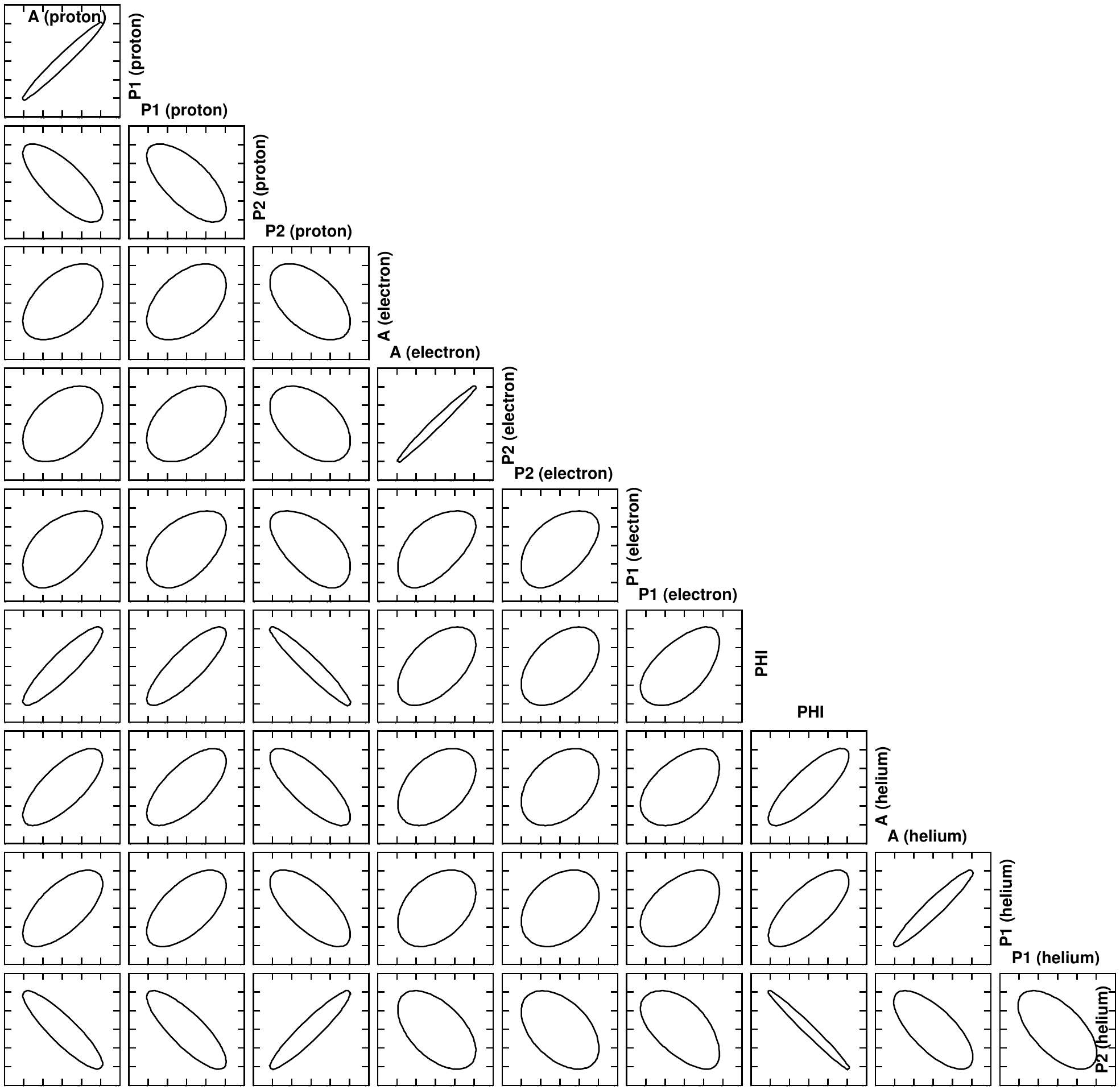}
\caption{Fit parameter error ellipses calculated at 1$\sigma$. For each plot we subtracted from the parameter its value given in Table \ref{table_resultat_fit} and divided by its uncertainty. In this graph the ellipses are therefore centered on 0; each axis is drawn from -1.5 to 1.5.}
\label{correlation}
\end{center}
\end{figure}

In Figure \ref{emissivity_contour} (bottom) we show the proton, Helium and electron-positron LIS resulting from the fit as well as their uncertainties. In the figure we also show the heliospheric fluxes used for the fit, and the electron-positron LIS. Folding those spectra with the production cross-section of Kamae {\it et al.} gives the $\gamma$-ray emissivity shown as a solid line in Figure \ref{emissivity_contour} (top) together with the 1$\sigma$ error contour. In this panel we decomposed the emissivity into a contribution from hadronic decay (red dash-dotted curve) and bremsstrahlung (green dash-double-dotted). We observe that a fit combining the heliospheric fluxes and the local emissivity under the hypothesis of the force-field approximation and of the LIS spectral form we chose reproduces the observations well. However the LAT emissivities at low energies seem systematically over-predicted while the high energy ones are under-predicted. The standard deviation corresponding to those variations is 14\% which is the same order as the difference of the three production cross-sections (Figure \ref{emissivity_cs}). We obtain an electron-positron LIS notably different from the AMS-01 observations.  However, retreiving an electron-positron LIS from the emissivity is challenging given that the bremsstrahlung contribution is significantly lower than the hadronic one and that the LAT emissivities at low energy might be biased by unaccounted for point sources given the broad LAT PSF below 100~MeV. We note that some of the 10 fit parameters are strongly correlated; we show in Figure \ref{correlation} the 2D error ellipses calculated at 1$\sigma$.

\begin{table}[h]
\caption{CR spectral form parameters deduced from the fit to a combination of the LAT emissivity, the proton and Helium fluxes from PAMELA and ATIC-2 and from the {\it Fermi} electron-positron observations.}
\begin{center}
\begin{tabular}{lccccccc}
    \hline
     &    & A   & P1 & P2  \\
     &    & m$^{-2}$ s$^{-1}$ sr$^{-1}$ GeV$^{-1}$ &  &   \\
    \hline
Proton      &    & (2.38 $\pm$ 0.09)$\times 10^4$   & 1.1 $\pm$ 0.6  &  2.838 $\pm$ 0.009   &\\
Helium      &    & (1.07 $\pm$ 0.02)$\times 10^3$  & 0.4 $\pm$ 0.3  &  2.777 $\pm$ 0.004  &\\
Electron-positron    &    & (3.8 $\pm$ 0.2)$\times 10^2$  & 4 $\pm$ 2 &  3.22 $\pm$ 0.02  &\\
\end{tabular}
\end{center}
\label{table_resultat_fit}
\end{table}

\begin{figure}
\begin{center}
\includegraphics[width=13cm,angle=270]{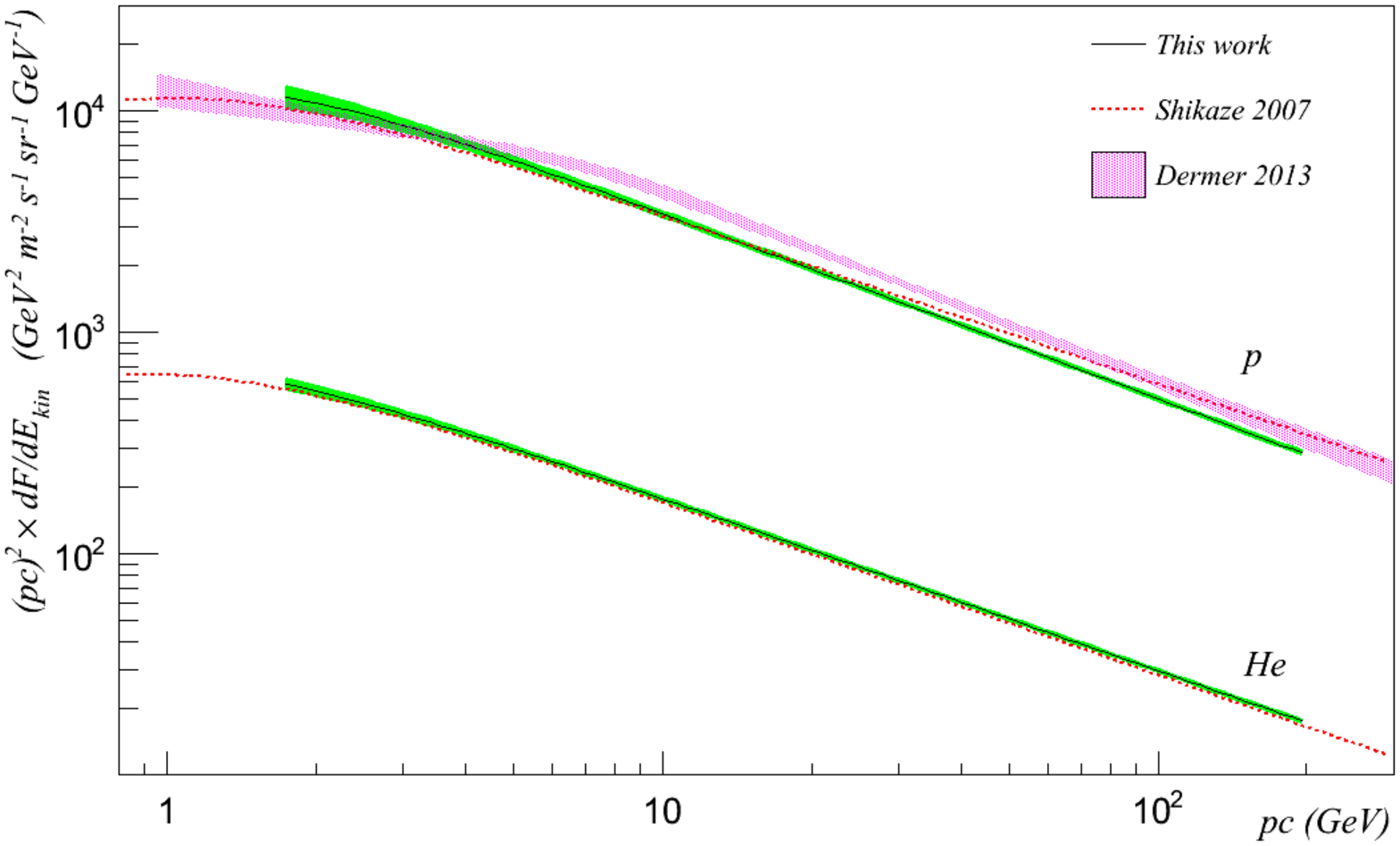}
\caption{Best fit proton and Helium spectral forms (black lines) versus the momentum per nucleon. The green contour corresponds to 1$\sigma$ uncertainty. The dashed line shows the LIS obtained by \cite{Shikaze:2007p3492} while the  magenta shaded area corresponds to that of \cite{Dermer:2013p4022}.}
\label{LIS_p_He_momentum}
\end{center}
\end{figure}

In Figure \ref{LIS_p_He_momentum} we display the proton and Helium LIS versus the CR momentum per nucleon. Although the best fit proton and Helium LIS show a hardening at low energy, we observe that within the uncertainties they are compatible with a power law in momentum. In the same graph we plotted the LIS obtained by \cite{Shikaze:2007p3492} and by \cite{Dermer:2013p4022}; the latter were derived from a preliminary set of the emissivities described here and with an updated version of the $\pi^{0}$ cross-section described in \cite{Kachelrie:2012p3635}.

\section{Summary}
From four years of {\it Fermi}-LAT observations we have derived the $\gamma$-ray emissivity per hydrogen atom between 50~MeV and 50~GeV in the solar neighborhood. We have measured its spectrum and estimated the systematic uncertainties, and we have shown that the spectral shape of the emissivity is the same within uncertainties for hydrogen in the atomic, molecular, ionized or DNM components of the ISM. We have extracted the molecular hydrogen-to-CO conversion factor  $X_{CO}=(0.902\pm0.007) \times 10^{20}$ cm$^{-2}$ (K km s$^{-1}$)$^{-1}$ and the dust-to-gas ratio $X_{DUST}=(41.4\pm0.3) \times 10^{20}$ cm$^{-2}$ mag$^{-1}$. We have shown that the WIM is not distributed like NE2001 prediction, nor does it follow a simple exponential scale-height. Only a small fraction of the WIM has the same morphology as the free-free emission. We did not observe any significant variation of the emissivity in four large sub-regions of the sky. We have compared the LAT emissivities with the ones predicted from various production cross-section models and heliospheric proton and Helium fluxes. We obtained a good agreement when those CR fluxes are solar-demodulated. We have fitted parametrized spectral forms to the CR heliospheric fluxes as well as to the LAT emissivities and deduced possible LIS for proton, Helium and electron-positron assuming a force-field approximation. The accuracy of $\gamma$-ray production cross-sections was a signficant limitation in deriving the CR spectra from the emissivities. We have also derived the solar modulation potential corresponding to the PAMELA observations.

The author would like to thank the members of the {\it Fermi}-LAT {\it Diffuse and Molecular Clouds} working group and particularly Andy W. Strong for their help to improve the manuscript.

The \textit{Fermi} LAT Collaboration acknowledges generous ongoing support
from a number of agencies and institutes that have supported both the
development and the operation of the LAT as well as scientific data analysis.
These include the National Aeronautics and Space Administration and the
Department of Energy in the United States, the Commissariat \`a l'Energie Atomique
and the Centre National de la Recherche Scientifique / Institut National de Physique
Nucl\'eaire et de Physique des Particules in France, the Agenzia Spaziale Italiana
and the Istituto Nazionale di Fisica Nucleare in Italy, the Ministry of Education,
Culture, Sports, Science and Technology (MEXT), High Energy Accelerator Research
Organization (KEK) and Japan Aerospace Exploration Agency (JAXA) in Japan, and
the K.~A.~Wallenberg Foundation, the Swedish Research Council and the
Swedish National Space Board in Sweden.
 
Additional support for science analysis during the operations phase is gratefully acknowledged from the Istituto Nazionale di Astrofisica in Italy and the Centre National d'\'Etudes Spatiales in France.

Some of the results in this paper have been derived using the HEALPix \citep{Gorski:2005p1076} package.

\bibliography{my_paper_local_emissivity}
\end{document}